\documentclass[oribibl]{llncs}

\usepackage{times}
\usepackage{latexsym}
\usepackage[mathscr]{euscript}
\usepackage{fullpage,amssymb,alltt,times,color}
\usepackage{url}
\usepackage{amsmath}
\usepackage{graphics}
\usepackage{proof}

\usepackage[
bookmarks=true,bookmarksnumbered=false,bookmarksopen=false,
breaklinks=true,pdfborder={0 0 0},backref=false,colorlinks=false]
{hyperref}

\allowdisplaybreaks

\usepackage{cmll}

\newcommand{\Proof}{\noindent {\sc Proof}. }

\newcommand{\lbd}{\lambda}

\newcommand{\set}[1]{\{#1\}}
\newcommand{\setc}[2]{\set{#1 \mid #2}}
\newcommand{\patc}[2]{\langle#1 \mid #2\rangle}

\newcommand{\seq}[2]{\infer{#2}{#1}}

\newcommand{\join}{\vee}		
\newcommand{\meet}{\wedge}		

\newcommand{\Alt}{ \,\mbox{\large\boldmath$\mid$}\,  }
\newcommand{\Sub}[3]{#1[#3/ #2]}
\newcommand{\Subi}[2]{#2/#1}
\newcommand{\Subm}[2]{#1[#2]}
\newcommand{\Submimp}[2]{#1\{#2\}}

\newcommand{\dl}{[\![} 			
\newcommand{\dr}{]\!]} 			

\newcommand{\lke}[4]{#1 \,|\, #2\vdash  #4}
\newcommand{\lkes}[4]{#1 \,|\, #2\vdash_S  #4}
\newcommand{\lkv}[4]{#1 \vdash #3 \,|\, #4}
\newcommand{\lkc}[4]{#1 \vdash  #4}

\newcommand{\lkE}[4]{#1 \,;\, #2\vdash  #4}
\newcommand{\lkV}[4]{#1 \vdash #3 \,;\, #4}

\newcommand{\inl}[1]{{\it inl}(#1)}
\newcommand{\inr}[1]{{\it inr}(#1)}
\newcommand{\fst}[1]{{\it fst}(#1)}
\newcommand{\snd}[1]{{\it snd}(#1)}
\newcommand{\fcase}[2]{\mbox{{\it case}}\; #1\; [#2]}
\newcommand{\fcasei}[2]{ #1\mapsto #2}
\newcommand{\suits}[2]{#1\,\bot\, #2}
\newcommand{\Vtov}[1]{#1^{\scriptscriptstyle\lozenge}}
\newcommand{\derel}[1]{#1^{\scriptscriptstyle\lozenge}}
\newcommand{\Vtoe}[1]{#1^{\scriptscriptstyle\blacklozenge}}
\newcommand{\Etoe}[1]{#1^{\scriptscriptstyle\lozenge}}

\newcommand{\padii}[1]{{{#1}^{\protect\rule{0pt}{0.4em}}}} 
\newcommand{\coupe}[2]{\padii{\left\langle {#1}\,\middle|\,{#2}\right\rangle }} 

\newcommand{\lbdtch}{\overline\lbd\mu\tilde\mu_T}
\newcommand{\lbdqch}{\overline\lbd\mu\tilde\mu_Q}

\newcommand{\notP}[1]{\neg^{\scriptscriptstyle +}{#1}}
\newcommand{\notN}[1]{\neg^{\scriptscriptstyle -}{#1}}

\newcommand{\CBVP}[1]{\dl#1\dr_{{\tt v}}^{\scriptscriptstyle +}}

\newcommand{\copaireP}[1]{[#1]}
\newcommand{\copaireC}[4]{[#1 \stackrel{#2,#3}{}#4]}

\newcommand{\lpar}{\parr}

\newcommand{\CPSP}[1]{#1_{\scriptscriptstyle{\it cps}}}

\newcommand{\Figbar}{{\center \rule{\hsize}{0.2mm}}} 

\newcommand{\reify}[1]{#1^{\ast}}

\newcommand{\Subwn}[2]{#1\{\!\!\{#2\}\!\!\}}
\newcommand{\Subimpl}[3]{#1\{#3/#2\}}

\newcommand{\LK}{\mathsf{LK}}
\newcommand{\LKQ}{\mathsf{LKQ}}

\newcommand{\contvartovar}[1]{k_{#1}}

\author{Pierre-Louis Curien (CNRS, Paris 7, and INRIA)
\and 
Guillaume Munch-Maccagnoni (Paris 7 and INRIA)}
\institute{}

\title{The duality of computation under focus}

\begin{document}
\maketitle
\begin{abstract}
We review the close relationship between  abstract machines for (call-by-name or call-by-value) $\lambda$-calculi (extended with Felleisen's $\cal C$) and sequent calculus, reintroducing on the way  Curien-Herbelin's  syntactic kit  expressing the duality of computation.
We use this kit to provide a term language for a presentation of $\LK$ (with conjunction, disjunction, and negation), and to transcribe cut elimination as (non confluent) rewriting.  A key slogan here, which may appear here in print for the first time, is that commutative cut elimination rules are explicit substitution propagation rules.
We then describe the focalised proof search discipline (in the classical setting), and narrow down the language and the rewriting rules to a confluent calculus 
(a variant of the second author's focalising system $\mathsf{L}$).
We then define a game of patterns and counterpatterns,  leading us  to a fully focalised finitary syntax for a synthetic presentation of classical logic, that provides a quotient on (focalised) proofs, abstracting out the order of decomposition of negative connectives.\footnote{A slighlty shorter version appears in the Proceedings of the Conference IFIP TCS, Brisbane, Sept. 2010, published as a Springer LNCS volume., With respect to the published conference version, the present version corrects some minor mistakes in the last section, and develops a bit further the material of Section 5.}

 \end{abstract}
 \section{Introduction} \label{introduction}
 This paper on one hand has  an expository purpose and on the other hand pushes further the syntactic investigations on the duality of computation undertaken in \cite{CH2000}.

Section \ref{intro-sec} discusses the relation between familiar {\em abstract machines} for the $\lambda$-calculus (extended with control) and (classical) {\em sequent calculus}.
Section  \ref{LK-proofs-sec} presents  a faithful language (with a one-to-one correspondence between well-typed terms and proof trees) for a presentation of LK that anticipates a key ingredient of focalisation, by choosing a dissymetric presentation for the conjunction on one side and the disjunction on the other side of sequents.  We recall the non-confluence of unconstrained classical cut-elimination.
 
In Section \ref{LKQ-sec}, we present the {\em focalised proof search discipline} (for  {\em classical logic}), and adapt the syntactic framework of Section \ref{LK-proofs-sec} to get a confluent
system whose normal forms are precisely the terms denoting  (cut-free) focalised proofs. The system we arrive at from these proof-search motivations  is (a variant of) the second author's {\em focalising system $\mathsf{L}$} ($\mathsf{L_{\textrm{foc}}}$) \cite{Munch2009} 
We prove the completeness of $\mathsf{L_{\textrm{foc}}}$ with respect to $\LK$ for provability.
 In Section \ref{encodings-sec}, we define some simple encodings having $\mathsf{L_{\textrm{foc}}}$ as source or target, indicating its suitability as an intermediate language (between languages and their execution or compilation).

 Finally, in Section \ref{LKQS-sec}, we further reinforce the focalisation discipline, which leads us to 
{\em synthetic system $\mathsf{L}$} ($\mathsf{L_{\textrm{synth}}}$),  a logic of synthetic connectives in the spirit of Girard's ludics and Zeilberger's CU, for which we offer a syntactic account based on a simple game of patterns and counterpatterns that can be seen as another manifestation of dualities of computation.  We show that the synthetic system $\mathsf{L}$ is complete with respect to focalising system $\mathsf{L}$.


\smallskip\noindent
{\em Notation.}   We shall write $\Subimpl{t}{x}{v}$ the result of substituting $v$ for $x$ at all (free) occurrences of $x$ in $t$,  and $\Sub{t}{x}{v}$ for an explicit  operator \cite{ACCL} added to the language together with rules propagating it.
Explicit substitutions are relevant here because they account for the commutative cut rules (see Section \ref{LK-proofs-sec}).

 \section{Abstract machines and sequent calculus} \label{intro-sec}
 In this section, we would like to convey
the idea that sequent calculus could have arisen from the goal of providing a typing system for
the states of an abstract machine for the 
``mechanical evaluation of expressions'' 
(to quote the title of Peter Landin's pioneering paper \cite{Landin64}).

Here is a simple device for executing a (closed) $\lbd$-term in call-by-name (Krivine machine \cite{KrivineMach}): 
\begin{center}
\fbox{$\begin{array}{lllllllll}
\coupe{MN}{E} & \longrightarrow&  \coupe{M}{N\cdot E} \quad &&& \quad
\coupe{\lbd x.M}{N\cdot E} & \longrightarrow &  \coupe{\Subimpl{M}{x}{N}}{E}
\end{array}$}
\end{center}
A state of the machine is thus a pair $\coupe{M}{E}$ where
 $M$ is ``where the computation is currently active'', and
 $E$ is the stack of  things that are waiting to be done in the future, or the continuation, or the evaluation context.
In $\lbd$-calculus litterature, contexts are more traditionally presented as terms with a hole: with this tradition,  
$\coupe{M}{E}$ (resp. $M\cdot E$) reads as $E[M]$ (resp. $E[[]M]$), or ``fill the hole of $E$  with $M$  (resp. $[]M$)''.

How can we type the components of this machine?
We have three categories of terms and of typing judgements:
$$\begin{array}{ccccccccc}
\mbox{Expressions} &&& \mbox{Contexts} &&& \mbox{Commands} 
\\
 M::= x\Alt \lbd x.M\Alt MM \quad&&&\quad
 E::= [\:]\Alt M\cdot E\quad &&& \quad
 c::=\coupe{M}{E}\\
 (\lkc{\Gamma}{}{}{M:A})\quad&&&\quad (\lke{\Gamma}{E:A}{}{R})
 \quad &&& \quad c:(\lkc{\Gamma}{}{}{R})
\end{array}$$
where $R$ is a (fixed) type of {\em final results}.  The type of an expression (resp. a context) is the type of the value that it is producing (resp. expecting).
The typing rules for  contexts and commands are as follows:
$$\seq{}{\lke{\Gamma}{ [\:]:R}{}{R}}\quad\quad
\seq{ \lkc{\Gamma}{}{}{ M:A}\quad \lke{\Gamma}{ E:B}{}{R}}
{\lke{\Gamma}{M\cdot E:A\rightarrow B}{}{R}}\quad\quad
\seq{\lkc{\Gamma}{}{}{ M:A}\quad\lke{\Gamma}{ E:A}{}{R}}
{\coupe{M}{E}:(\lkc{\Gamma}{}{}{R})}$$
and the typing rules for expressions are the usual ones for simply typed $\lambda$-calculus.
Stripping up the term information, the second and third rules are rules of
{\em sequent calculus} (left introduction of implication and cut).

We next review  Griffin's typing of  Felleisen's control operator ${\cal  C}$. 
As a matter of fact, the behaviour of this  constructor is best expressed at the level of an abstract machine:
\begin{center}
\fbox{$\begin{array}{lllllllll}
\coupe{{\cal C}(M)}{E} & \longrightarrow &\coupe{M}{\reify{E}\cdot[\:]} \quad &&& \quad
\coupe{\reify{E}}{N\cdot E'} &\longrightarrow& \coupe{N}{E}
\end{array}$}
\end{center}
The first rule explains how the  continuation $E$ gets {\em captured}, and the second rule how it gets {\em restored}.
 Griffin \cite{Griffin90} observed  that the typing constraints induced by the well-typing  of these four commands are
met when  ${\cal C}(M)$  and $\reify{E}$ are typed as follows:
$$\begin{array}{cccc}
\seq{\lkc{\Gamma}{}{}{M: (A\rightarrow R)\rightarrow R}}
{\lkc{\Gamma}{}{}{{\cal C}(M): A}} \quad&&& \quad\seq{\lke{\Gamma}{E: A}{}{R}}{\lkc{\Gamma}{}{}{\reify{E}: A\rightarrow R}}
\end{array}$$
These are the rules that one adds to intutionistic natural deduction to make it classical, if we interpret $R$ as  $\bot$ (false), and if we encode $\neg A$ as $A\rightarrow R$.
Hence, Griffin  got no less than {Curry-Howard for classical logic! But how does this sound in sequent calculus style?
In classical sequent calculus, sequents have  several formulas on the right and
$\lkc{\Gamma}{}{}{\Delta}$ reads as ``if all formulas in $\Gamma$ hold, then at least one formula of $\Delta$ holds''.  Then it is natural to associate  continuation variables with the formulas in $\Delta$: a term will depend on its input variables, and on its output continuations.
With this in mind, we can read  the operational rule for ${\cal C}(M)$ as
`` ${\cal C}(M)$ is  a map $E \mapsto \coupe{M}{\reify{E}\cdot[\:]}$'',
and write it with a new binder (that comes from \cite{Parigot92}):
$${\cal C}(M) =\mu\beta.\coupe{M}{\reify{\beta}\cdot[\:]}$$
where $[\:]$ is now a continuation variable (of  ``top-level'' type $R$).
Likewise, we synthesise $\reify{E}=\lbd x.\mu\alpha.\coupe{x}{E}$, with $\alpha,x$ fresh, from the operational rules for $\reify{E}$ and for $\lambda x.M$.

The typing judgements are now:
$
 (\lkv{\Gamma}{}{M:A}{\Delta})$, 
$(\lke{\Gamma}{E:A}{}{\Delta})$, and
$c:(\lkc{\Gamma}{}{}{\Delta})
$.
The two relevant new typing rules are (axiom, right {\em activation}):
\begin{center}
$\begin{array}{llll}
\seq{}{\lke{\Gamma}{\alpha:A}{}{\alpha:A,\Delta}} &&&
\seq{c:(\lkc{\Gamma}{}{}{\alpha:A,\Delta})}{ \lkv{\Gamma}{}{\mu\alpha.c:A}{\Delta}}
\end{array}$
\end{center}
plus a reduction rule: $\coupe{\mu\alpha.c}{E}\longrightarrow\Subimpl{c}{\alpha}{E}$. 

Note that in this setting, there is no more need to ``reify'' a context  $E$ into an expression $\reify{E}$, as it can be directly substituted for a continuation variable.

Similarly, we can read off a (call-by-name) definition of $MN$ from  its operational rule:
 $MN=\mu\beta.\coupe{M}{N.\beta}$. Hence we can
 remove application from the syntax and arrive at a system in sequent calculus style {\em only}} (no more elimination rule).
This yields Herbelin's $\overline{\lambda}\mu$-calculus \cite{Her95}:
$$
\mbox{Expressions} \;\;  M::= x\Alt \lbd x.M\Alt \mu\alpha.c\quad\quad
\mbox{Contexts} \;\;  E::= \alpha\Alt M\cdot E \quad\quad
\mbox{Commands} \;\; c::=\coupe{M}{E} 
$$
 which combines the first two milestones above: ``sequent calculus'', ``classical".

\medskip
Let us step back to the $\lbd$-calculus. The following describes a call-by-value version of Krivine machine:
\begin{center}
\fbox{$\begin{array}{llllllll}
\coupe{MN}{e} & \longrightarrow &  \coupe{N}{M\odot e} \quad&&&\quad
\coupe{V}{M\odot e}& \longrightarrow &  \coupe{M}{V\cdot e}
\end{array}$}
\end{center}
(the operational rule for $\lambda x.M$ is unchanged)\footnote{The reason for switching notation  from $E$ to $e$ will become clear in Section \ref{encodings-sec}.}. 
Here, $V$ is a {\em value}, defined as being either a variable or an abstraction (this goes back to \cite{PlotkinCBNV}).
Again, we can read  $M\odot e$ as  
``a map $V\mapsto \coupe{M}{V\cdot e}$'', 
or, introducing a new binder $\tilde\mu$ (binding now ordinary variables):
$$M\odot e = \tilde\mu x. \coupe{M}{x\cdot e}$$
The typing rule for this operator is (left activation):
\begin{center}
$\begin{array}{c}
\seq{ c:(\lkc{\Gamma,x:A}{}{}{\Delta})}{\lke{\Gamma}{\tilde\mu x.c:A}{}{\Delta}}
\end{array}$
\end{center}
and the operational rule is
$\coupe{V}{\tilde\mu x.c}\longrightarrow \Subimpl{c}{x}{V}\;\;(V\mbox{ value})$.

\smallskip
Finally, we get from the rule for $MN$ a  call-by-value definition of application:
$MN=\mu\alpha.\coupe{N}{\tilde\mu x.\coupe{M}{x\cdot \alpha}}$.

\smallskip
We have arrived at Curien and Herbelin's  $\lbdqch$-calculus  \cite{CH2000}:
\begin{center}
\begin{tabular}{|cc|cc|cc|c|}
\hline
Expressions  $ M::= \Vtov{V}\Alt \mu\alpha.c $ && \:Values $\; V::= x\Alt \lbd x.M$ && \:Contexts $\; e::= \alpha\Alt V\cdot e\Alt \tilde\mu x.c$ && \:Commands $\;  c::=\coupe{M}{e}$\\
 $ \lkv{\Gamma}{}{M:A}{\Delta}$  && $\lkV{\Gamma}{}{V:A}{\Delta}$ &&
 $\lke{\Gamma}{e:A}{}{\Delta} $
 &&
 $c:(\lkc{\Gamma}{}{}{\Delta})$\\
 \hline
\end{tabular}
\end{center}
with a new judgement for values (more on this later) and an
 explicit coercion from values to expressions. The syntax for contexts  is both extended ($\tilde\mu x.c$) and restricted ($V\cdot e$ instead of $M\cdot e$). The reduction rules are as follows:
\begin{center}
\fbox{$\begin{array}{lllllllll}
\coupe{\Vtov{(\lbd x.M)}}{V\cdot e}\longrightarrow \coupe{\Subimpl{M}{x}{V}}{e} \quad&&\quad
\coupe{\mu\alpha.c}{e}\longrightarrow \Subimpl{c}{\alpha}{e}\quad &&\quad
\coupe{\Vtov{V}}{\tilde\mu x.c}\longrightarrow \Subimpl{c}{x}{V}
\end{array}$}
\end{center}
\section{A language for $\LK$ proofs} \label{LK-proofs-sec}
In this section, we use some of the kit of the previous section to give a term language for classical sequent calculus $\LK$, with negation, conjunction, and disjunction as connectives.  Our term language is as follows:
$$\begin{array}{lllll}
\mbox{Commands} &&&& c::=\coupe{x}{\alpha} \Alt \coupe{v}{\alpha} \Alt \coupe{x}{e} \Alt \coupe{\mu\alpha.c}{\tilde\mu x.c}\\
\mbox{Expressions} &&&& v::= (\tilde\mu x.c)^\bullet \Alt (\mu\alpha.c,\mu\alpha.c)\Alt \inl{\mu\alpha.c}\Alt \inr{\mu\alpha.c}\\
\mbox{Contexts} &&&& e::= \tilde\mu\alpha^\bullet.c \Alt \tilde\mu (x_1,x_2).c \Alt  \tilde\mu\copaireP{\inl{x_1}.c_1|\inr{x_2}.c_2}
\end{array}$$
(In $\coupe{v}{\alpha}$ (resp. $\coupe{x}{e}$), we suppose $\alpha$ (resp. $x$) fresh for  $v$ (resp. $e$).) A {\em term} $t$ is a command, an expression, or a context.
As in section \ref{intro-sec}, we have three kinds of sequents:
$(\Gamma\vdash\Delta)$, 
$(\lkv{\Gamma}{}{A}{\Delta})$, and $(\lke{\Gamma}{A}{}{\Delta})$.
We decorate $\LK$'s inference rules with terms,  yielding the following typing system (one term construction for each rule of $\LK$):
 \begin{center}
{\small (axiom and cut/contraction) $\quad\quad \seq{}{\coupe{x}{\alpha}:(\Gamma,x:A\vdash \alpha:A,\Delta)}\quad\quad
\seq{c:(\Gamma\vdash \alpha: A,\Delta)\quad\quad
d:(\Gamma, x:A\vdash\Delta)}
{\coupe{\mu\alpha.c}{\tilde\mu x.d}:(\Gamma\vdash\Delta)}$

\bigskip
$(\mbox{right})\quad\quad\quad\quad\quad\seq{c:(\Gamma,x:A\vdash\Delta)}{\lkv{\Gamma}{}{(\tilde\mu x.c)^\bullet:\neg A}{\Delta}}\quad\quad\seq{c_1:(\Gamma\vdash \alpha_1:A_1,\Delta) \quad\quad c_2:(\Gamma\vdash \alpha_2:A_2,\Delta)}
{\lkv{\Gamma}{}{(\mu\alpha_1.c_1,\mu\alpha_2.c_2):A_1\wedge A_2}{\Delta}}$

\medskip
$
\seq{c_1:(\Gamma\vdash \alpha_1:A_1,\Delta)}{\lkv{\Gamma}{}{\inl{\mu\alpha_1.c_1}:A_1\vee A_2}{\Delta}}\quad\quad
\seq{c_2:(\Gamma\vdash \alpha_2:A_2,\Delta)}{\lkv{\Gamma}{}{\inr{\mu\alpha_2.c_2}:A_1\vee A_2}{\Delta}}$

\bigskip
$(\mbox{left})\quad \seq{c:(\Gamma\vdash \alpha:A,\Delta)}{\lke{\Gamma}{\tilde\mu\alpha^\bullet.c:\neg A}{}{\Delta}}\quad\quad
\seq{c:(\Gamma,x_1:A_1,x_2:A_2\vdash\Delta)}{\lke{\Gamma}{\tilde\mu(x_1,x_2).c:A_1\wedge A_2}{}{\Delta}}\quad\quad\seq{c_1:(\Gamma,x_1:A_1\vdash\Delta)\quad\quad c_2:(\Gamma,x_2:A_2\vdash\Delta)}
{\lke{\Gamma}{\tilde\mu\copaireP{\inl{x_1}.c_1|\inr{x_2}.c_2}:A_1\vee A_2}{}{\Delta}}$

\bigskip
$(\mbox{deactivation})\quad\quad\seq{\lkv{\Gamma}{}{v:A}{\Delta}}{\coupe{v}{\alpha}:(\Gamma\vdash\alpha:A,\Delta)}\quad\quad
\quad\quad\seq{\lke{\Gamma}{e:A}{}{\Delta}}{\coupe{x}{e}:(\Gamma,x:A\vdash\Delta)}$}
\end{center}
Note that the activation rules are packaged in the introduction rules and in the cut rule. As for the underlying sequent calculus rules, we have made the following choices:
\begin{enumerate}
\item We have preferred {\em additive} formulations for the  cut rule and for the right introduction of conjunction (to stay in tune with the tradition of typed $\lambda$-calculi) over a multiplicative one
where the three occurrences of $\Gamma$ would be resp. $\Gamma_1$, $\Gamma_2$, and $\Gamma_1,\Gamma_2$ (idem for $\Delta$).  An important consequence of this choice is that contraction is a derived rule of our system, whence the name of {\em cut/contraction} rule above\footnote{In usual syntactic accounts of contraction, one says that if,  say  $t$ denotes a proof of $\Gamma,x:A,y:A\vdash\Delta$, then $\Subm{t}{\Subi{x}{z},\Subi{y}{z}}$ denotes a proof of $\Gamma,z:A\vdash\Delta$. Note that if this substitution is explicit, then we are back to an overloading of cut and contraction.}:
\begin{center}
$\seq{\seq{}{\lkc{\Gamma,A}{}{}{A,\Delta}}\quad\quad \lkc{\Gamma,A,A}{}{}{\Delta}}{\lkc{\Gamma,A}{}{}{\Delta}}\quad\quad\quad\seq{\lkc{\Gamma}{}{}{A,A,\Delta}\quad\quad \seq{}{\lkc{\Gamma,A}{}{}{A,\Delta}}}{\lkc{\Gamma}{}{}{A,\Delta}}$
\end{center}
\item Still in the $\lambda$-calculus tradition, weakening  is ``transparent''. If $c:\Gamma\vdash\Delta$ is well-typed, then  $c:(\Gamma,\Gamma'\vdash\Delta,\Delta')$  is well-typed (idem $v,e$).
(Also, we recall that all free variables of $c$ are among the ones declared in $\Gamma,\Delta$.) 
\item More importantly, we have adopted {\em irreversible} rules for right introduction of disjunction. On the other hand, we have given  a {\em reversible} rule for left introduction of conjunction: the
premise is derivable from the conclusion. {\em This choice  prepares the ground for the next section on focalisation.}\footnote{For the same reason, we have chosen to take three connectives instead of just two, say, $\join$ and $\neg$, because in the focalised setting 
$\neg(\neg A\join \neg B)$ is only equivalent to $A\meet B$ at the level of {\em provability}.}
\end{enumerate}
The relation between our typed terms and $\LK$ proofs is as follows.

\smallskip\noindent
-  Every typing proof  induces a proof tree of $\LK$ (one erases variables naming assumptions and conclusions, terms, the distinction between the three kinds of sequents, and the application of the deactivation rules). 

\smallskip\noindent
- If bound variables are explicitly typed  (which we shall refrain from doing in the sequel), 
then
every provable  typing judgement, say $\lke{\Gamma}{e:A}{}{\Delta}$,
has a unique  typing proof, i.e. all  information  is in $\Gamma$, $A$, $\Delta$, $e$.

\smallskip\noindent
- If $\Pi$ is  an $\LK$ proof tree  of $(A_1,\ldots,A_m\vdash B_1,\ldots,B_n)$, and if names
$x_1,\ldots,x_m$, $\alpha_1,\ldots,\alpha_n$ are provided, then there exists  a unique command $c:(\lkc{x_1:A_1,\ldots,x_m:A_m}{}{}{\alpha_1:B_1,\ldots,\alpha_n:B_n})$, whose (unique) typing proof gives back $\Pi$ by erasing.

\smallskip
With this syntax, we can express the cut-elimination rules of $\LK$ as {\em rewriting rules}:


\medskip
Logical rules (redexes of the form $\coupe{\mu\alpha.\coupe{v}{\alpha}}{\tilde\mu x.\coupe{x}{e}}$):
\begin{center}
$\begin{array}{l}
\coupe{\mu\alpha.\coupe{(\tilde\mu x.c)^\bullet}{\alpha}}
{\tilde\mu y.\coupe{y}{\tilde\mu\alpha^\bullet.d}}
\longrightarrow  \coupe{\mu\alpha.d}{\tilde\mu x.c}\quad\quad
(\mbox{similar rules for  conjunction and disjunction})
\end{array}$
\end{center}

Commutative rules (going ``up left'', 
redexes  of the form $\coupe{\mu\alpha.\coupe{v}{\beta}}{\tilde\mu x.c}$):
$$\begin{array}{l}
\coupe{\mu\alpha.\coupe{(\tilde\mu y.c)^\bullet}{\beta}}{\tilde\mu x.d} \longrightarrow \coupe{\mu\beta'.\coupe{(\tilde\mu y.\coupe{\mu\alpha.c}{\tilde\mu x.d})^\bullet}{\beta'}}{\tilde\mu y.\coupe{y}{\beta}} \quad(\neg\mbox{ right})\\
(\mbox{similar rules of commutation with the other right introduction rules and with the left introduction rules})\\

\coupe{\mu\alpha.\coupe{\mu\beta.\coupe{y}{\beta}}{\tilde\mu y'.c}}{\tilde\mu x.d}\longrightarrow \coupe{\mu\beta.\coupe{y}{\beta}}{\tilde\mu y'.\coupe{\mu\alpha.c}{\tilde\mu x.d}} \quad(\mbox{contraction right})\\
\coupe{\mu\alpha.\coupe{\mu\beta'.c}{\tilde\mu y.\coupe{y}{\beta}}}{\tilde\mu x.d} \longrightarrow \coupe{\mu\beta'.\coupe{\mu\alpha.c}{\tilde\mu x.d}}{\tilde\mu y.\coupe{y}{\beta}} \quad (\mbox{contraction left})\\
\coupe{\mu\alpha.\coupe{\mu\alpha'.c}{\tilde\mu x'.\coupe{x'}{\alpha}}}{\tilde\mu x.d} \longrightarrow \coupe{\mu\alpha.\coupe{\mu\alpha'.c}{\tilde\mu x.d}}{\tilde\mu x.d} \quad(\mbox{duplication})\\
\coupe{\mu\alpha.\coupe{y}{\beta}}{\tilde\mu x.d} \longrightarrow \coupe{y}{\beta} \quad (\mbox{erasing})
\end{array}$$

Commutative rules (going ``up right'', redexes of the form $\coupe{\mu\alpha.c}{\tilde\mu x.\coupe{y}{e}}$ ):
similar rules.

\smallskip
The (only?) merit of this syntax is its tight fit with proof trees and traditional cut elimination defined as transformations of undecorated proof trees. If we accept to losen this,
we arrive at the following  more ``atomic'' syntax:
$$\begin{array}{lllll}
\mbox{Commands} &&&& c::=\coupe{v}{e} \Alt \Subm{c}{\sigma}\\
\mbox{Expressions} &&&& v::= x\Alt \mu\alpha.c \Alt e^\bullet \Alt (v,v)\Alt \inl{v}\Alt \inr{v}\Alt \Subm{v}{\sigma}\\
\mbox{Contexts} &&&& e::= \alpha\Alt \tilde\mu x.c\Alt \tilde\mu\alpha^\bullet.c \Alt \tilde\mu (x_1,x_2).c \Alt  \tilde\mu\copaireP{\inl{x_1}.c_1|\inr{x_2}.c_2}\Alt \Subm{e}{\sigma}
\end{array}$$
where $\sigma$ is a list $\Subi{x_1}{v_1},\ldots,\Subi{x_m}{v_m},\Subi{\alpha_1}{e_1},\ldots,
\Subi{\alpha_n}{e_n}$.
In this syntax, activation becomes ``first class'', and two versions of the axiom are now present ($x$, $\alpha$, which give back the axiom of the previous syntax by deactivation). 
The typing rules are as follows (we omit the rules for $\tilde\mu x.c$, $\tilde\mu\alpha^\bullet.c$, $\tilde\mu (x_1,x_2).c$,  $\tilde\mu\copaireP{\inl{x_1}.c_1|\inr{x_2}.c_2}$, which are unchanged):
{\small $$\seq{}{\lkv{\Gamma\:,\:x:A}{}{x:A}{\Delta}}\quad\quad
\seq{}{\lke{\Gamma}{\alpha:A}{}{\alpha:A\:,\:\Delta}}\quad\quad
\seq{\lkv{\Gamma}{}{v:A}{\Delta}\quad\quad
\lke{\Gamma}{e:A}{}{\Delta}}
{\coupe{v}{e}:(\lkc{\Gamma}{}{}{\Delta})}
$$
$$
\seq{c:(\lkc{\Gamma\:,\:x:A}{}{}{\Delta})}{\lke{\Gamma}{\tilde\mu x.c:A}{}{\Delta}}
\quad\quad \seq{c:(\lkc{\Gamma}{}{}{\alpha:A\:,\:\Delta})}{\lkv{\Gamma}{}{\mu\alpha.c:A}{\Delta}}
$$
$$\seq{\lke{\Gamma}{e:A}{}{\Delta}}{\lkv{\Gamma}{}{e^\bullet:\neg A}{\Delta}}\quad\quad\seq{\lkv{\Gamma}{}{v_1:A_1}{\Delta} \quad\quad \lkv{\Gamma}{}{v_2:A_2}{\Delta}}
{\lkv{\Gamma}{}{(v_1,v_2):A_1\meet A_2}{\Delta}}\quad\quad
\seq{\lkv{\Gamma}{}{v_1:A_1}{\Delta}}{\lkv{\Gamma}{}{\inl{v_1}:A_1\join A_2}{\Delta}}\quad\quad
\seq{\lkv{\Gamma}{}{v_2:A_2}{\Delta}}{\lkv{\Gamma}{}{\inr{v_2}:A_1\join A_2}{\Delta}}$$
$$\seq{c:(\lkc{\Gamma,x_1:A_1,\ldots,x_m:A_m}{}{}{\alpha_1:B_1,\ldots,\alpha_n:B_n})\;\ldots\;\lkv{\Gamma}{}{v_i:A_i}{\Delta}\;\ldots\;\ldots\;\lke{\Gamma}{e_j:B_j}{}{\Delta}\;\ldots}{\Subm{c}{\Subi{x_1}{v_1},\ldots,\Subi{x_m}{v_m},\Subi{\alpha_1}{e_1},\ldots,
\Subi{\alpha_n}{e_n}}:(\lkc{\Gamma}{}{}{\Delta})} \quad
(\mbox{idem } \Subm{v}{\sigma}, \Subm{e}{\sigma})$$}

Note that we also have now
{\em explicit substitutions}
  $\Subm{t}{\sigma}$, which feature a form of (multi-)cut where the receiver $t$'s active formula, if any, is not among the cut formulas, in contrast with the construct $\coupe{v}{e}$ where the cut formula is active on both sides.

It is still the case that, by erasing, a well-typed term of this new syntax induces a proof of $\LK$, and that all proofs of $\LK$ are reached (although not injectively anymore), since all terms of the previous syntax are terms of the new syntax. 
The rewriting rules divide now  in {\em three} groups:
$$\begin{array}{lllll}
(\mbox{control}) && \coupe{\mu\alpha.c}{e}  \longrightarrow  \Sub{c}{\alpha}{e} \quad\quad\quad\quad\quad\quad
 \coupe{v}{\tilde\mu x.c}  \longrightarrow  \Sub{c}{x}{v}\\

(\mbox{logical}) &&\coupe{e^\bullet}{\tilde\mu \alpha^\bullet.c}  \longrightarrow  \Sub{c}{\alpha}{e} \quad\quad\quad\quad\quad \coupe{(v_1,v_2)}{\tilde\mu(x_1,x_2).c} \longrightarrow  \Subm{c}{\Subi{x_1}{v_1},\Subi{x_2}{v_2}}\\
&& \coupe{\inl{v_1}}{\tilde\mu\copaireP{\inl{x_1}.c_1|\inr{x_2}.c_2}}  \longrightarrow  \Sub{c_1}{x_1}{v_1}\quad
(\mbox{idem }{\it inr})\\
(\mbox{commutation}) &&
\Subm{\coupe{v}{e}}{\sigma}\longrightarrow \coupe{\Subm{v}{\sigma}}{\Subm{e}{\sigma}}\\
&&
\Subm{x}{\sigma}\longrightarrow x \;\; (x\mbox{ not declared in }\sigma)\quad\quad
\Subm{x}{\Subi{x}{v},\sigma}\longrightarrow v    \quad (\mbox{idem }\Subm{\alpha}{\sigma})\\
&& \Subm{(\mu\alpha.c)}{\sigma} \longrightarrow \mu\alpha.(\Subm{c}{\sigma})   \quad(\mbox{idem } \Subm{(\tilde\mu x.c)}{\sigma}) \quad(\mbox{capture avoiding})\\
&& (\mbox{etc, no rule for composing substitutions})
\end{array}$$
The {\em control rules} mark the decision to launch a substitution (and, in this section, of the direction in which to go, see below).   The {\em logical rules} provide
the interesting cases of cut elimination, corresponding to cuts where the active formula has been just introduced on both sides. 
The {\em commutative cuts} are now accounted for ``trivially''
by means of the {\em explicit substitution machinery}
that carries substitution progressively inside terms towards their variable occurrences.
Summarising, by liberalising the syntax, we have gained
considerably in readability of the cut elimination rules\footnote{The precise relation with the previous rules is as follows:
for all $s_1,s_2$ such that  $s_1\longrightarrow s_2$ in the first system,  there exists $s$  such that $s_1\longrightarrow^* s {}^*\longleftarrow s_2$  in the new system, e.g.,
 for  ($\neg$ right)
$\coupe{\mu\alpha.\coupe{(\tilde\mu y.c)^\bullet}{\beta}}{e}$  
$\longrightarrow^*$
$\coupe{(\tilde\mu y.(\Sub{c}{\alpha}{e}))^\bullet}{\beta}$
$ {}^*\!\!\longleftarrow$
$\coupe{\mu\beta'.\coupe{(\tilde\mu y.\coupe{\mu\alpha.c}{e})^\bullet}{\beta'}}{\tilde\mu y.\coupe{y}{\beta}}$.}.

\begin{remark}  
In the ``atomic''  syntax, contractions are transcribed as terms of the form $\coupe{v}{\beta}$ where $\beta$ occurs free in $v$, or of the form $\coupe{x}{e}$ where $x$ occurs freely in $e$.
If $\beta$  (resp. $x$) does not occur free in $v$ (resp. $e$), then the command expresses a simple deactivation.
\end{remark}

The problem with classical logic viewed as a computational system is its wild non confluence, 
as captured by Lafont's critical pair \cite{Girard89,DJSDec}, for which the $\mu\tilde\mu$ kit offers a crisp formulation.  For any $c_1,c_2$ both of type $(\Gamma\vdash\Delta)$,  we have (with $\alpha,x$ fresh for $c_1,c_2$, respectively):
\begin{center}
$ c_1\quad{}^*\!\longleftarrow \quad \coupe{\mu\alpha.c_1}{\tilde\mu x.c_2}\quad\longrightarrow^*\quad c_2$
\end{center}
So, all proofs are identified... {\em Focalisation}, discussed in the next section, will  guide us to solve this dilemma.
 \section{A syntax for focalised classical logic} \label{LKQ-sec}
In this section, we adapt the
  {\em focalisation discipline} (originally introduced by \cite{Andreoli92} in the setting of linear logic) to $\LK$.
  A focalised proof search
 alternates between right and left phases, as follows:
  
\smallskip\noindent
-  {\it Left phase}: Decompose (copies of) formulas on the left, in any order. Every decomposition of a negation on the left feeds the right part of the sequent. At any moment, one can change the phase from left to right.

\smallskip\noindent
- {\it Right phase}:  Choose a formula $A$ on the right, and
{\em hereditarily} decompose a copy of  it in all branches of the proof search. This {\em focusing} in any branch can only end with an axiom (which ends the proof search in that branch), or with a decomposition of a negation, which prompts a phase change back to the left. Etc\ldots

\smallskip
Note the irreversible (or {\em positive, active}) character of the whole right phase, by the choice of $A$, by the choice of the left or right summand of a disjunction. One takes the risk of not being able to eventually end a proof search branch with an axiom. In contrast, all the choices on the left are reversible (or {\em negative, passive}).
This strategy is not only complete (see below), it also guides us to design a disciplined logic whose behaviour will not collapse all the proofs.

\smallskip
To account for right focalisation, we introduce a fourth kind of judgement and a fourth syntactic category of terms: the {\em values}, typed as $(\lkV{\Gamma}{}{V:A}{\Delta})$ (the zone between the turnstyle and the semicolon is called the {\em stoup}, after \cite{GirardLC}). We also make official the existence of two disjunctions (since the behaviours of the conjunction on the left and of the disjunction on the right are different) and two conjunctions, by renaming $\meet,\join,\neg$ as $\otimes,\oplus,\notP{}$, respectively. Of course, this choice of linear logic like notation is not fortuitous. Note however that the source of distinction is not based here on the use of resources like in the founding work on linear logic, which divides the line between {\em additive} and {\em multiplicative} connectives. In contrast, our motivating dividing line here is that between {\em irreversible} and {\em reversible} connectives, and hopefully this provides additional motivation for the two conjunctions and the two disjunctions.
Our formulas are thus defined by the following syntax:
\begin{center}
$P::=X\Alt P\otimes P\Alt P\oplus P\Alt \notP{P}$
\end{center}
These formulas are called positive.
We can define their De Morgan duals as follows:
\begin{center}
$\overline{P_1\otimes P_2}=\overline{P_1}\lpar\overline{P_2}\quad\quad
\overline{P_1\oplus P_2}=\overline{P_1}\with\overline{P_2}\quad\quad
\overline{\notP{P}}=\notN{\overline{P}}$
\end{center}
These duals are {\em negative} formulas:
$N::=\overline{X} \Alt N\lpar N\Alt N\with N\Alt \notN{N}$. They restore the duality of connectives, and are implicit in the presentation that follows (think of $P$ on the left as being
a $\overline{P}$ in a unilateral sequent $\vdash \overline{\Gamma},\Delta$).

We are now ready to give the syntax of our calculus, which is a variant of the one given by the second author in \cite{Munch2009}\footnote{The main differences with  the system presented in  \cite{Munch2009} is that we have here
an explicit syntax of values, with an associated form of typing judgement, while focalisation is dealt with at the level of the reduction semantics in \cite{Munch2009} (see also Remark \ref{LK-LKQ-not-red-refl}).
Also, the present system is bilateral but limited to positive formulas on both sides, it thus corresponds to the positive 
subsystem
of the bilateral version of $\mathsf{L}_{\textrm{foc}}$ as presented in \cite{Munch2009}[long version, Appendix A].}. 

\begin{center}
\fbox{$\begin{array}{lllllll}
\mbox{Commands} && c::=\coupe{v}{e} \Alt \Subm{c}{\sigma}&&\\
\mbox{Expressions} && v::= \Vtov{V}\Alt \mu\alpha.c \Alt \Subm{v}{\sigma}\\
\mbox{Values} && V::= x \Alt  (V,V) \Alt \inl{V}\Alt \inr{V}\Alt e^\bullet \Alt \Subm{V}{\sigma}&& 
\\
\mbox{Contexts} && e::=\alpha \Alt \tilde\mu x.c\Alt  \tilde\mu\alpha^\bullet.c \Alt \tilde\mu (x_1,x_2).c \Alt  \tilde\mu\copaireP{\inl{x_1}.c_1|\inr{x_2}.c_2}\Alt \Subm{e}{\sigma}&& 
\end{array}$}
\end{center}
The typing rules are given in  Figure 1. Henceforth, we shall refer to the calculus of this section (syntax + rewriting rules) as $\mathsf{L}_{\textrm{foc}}$, and to the typing system as $\LKQ$  (after
\cite{DJSDec}).
Here are examples of proof terms in $\LKQ$.
\begin{example} \label{LKQ-proofs-ex}
\begin{itemize}
\item[]
$(\lkV{}{}{(\tilde\mu(x,\alpha^\bullet).\coupe{\Vtov{x}}{\alpha})^\bullet:\notP{(P\otimes \notP{P})}}{}$), where $\tilde\mu(x,\alpha^\bullet).c$ is an abbreviation for $\tilde\mu(x,y).\coupe{\Vtov{y}}{\tilde\mu\alpha^\bullet.c}$.

\item[]
 $\coupe{\Vtov{\inr{(\tilde\mu x.\coupe{\Vtov{\inl{x}}}{\alpha})^\bullet}}}{\alpha}:(\lkc{}{}{}{\alpha:P\oplus\notP{P}})$.
\item[]
 $(\lke{}{\tilde\mu(x_2,x_1).\coupe{\Vtov{(x_1,x_2)}}{\alpha}:P_2\otimes P_1}{}{\alpha:P_1\otimes P_2}$).
\end{itemize}
\end{example}
\begin{figure} \label{LKQ-fig}
\caption{System $\LKQ$}
\begin{center}
\fbox{$\seq{}{\lkV{\Gamma\:,\:x:P}{}{x:P}{\Delta}}\quad\quad
\seq{}{\lke{\Gamma}{\alpha:P}{}{\alpha:P\:,\:\Delta}}\quad\quad
\seq{\lkv{\Gamma}{}{v:P}{\Delta}\quad\quad
\lke{\Gamma}{e:P}{}{\Delta}}
{\coupe{v}{e}:(\lkc{\Gamma}{}{}{\Delta})}
$}

\medskip
\fbox{$
\seq{c:(\lkc{\Gamma\:,\:x:P}{}{}{\Delta})}{\lke{\Gamma}{\tilde\mu x.c:P}{}{\Delta}}
\quad\quad \seq{c:(\lkc{\Gamma}{}{}{\alpha:P\:,\:\Delta})}{\lkv{\Gamma}{}{\mu\alpha.c:P}{\Delta}}\quad\quad \seq{\lkV{\Gamma}{}{V:P}{\Delta}}{\lkv{\Gamma}{}{\Vtov{V}:P}{\Delta}}
$}

\medskip
\fbox{$\seq{\lke{\Gamma}{e:P}{}{\Delta}}{\lkV{\Gamma}{}{e^\bullet:\notP{P}}{\Delta}}\quad\quad\seq{\lkV{\Gamma}{}{V_1:P_1}{\Delta} \quad\quad \lkV{\Gamma}{}{V_2:P_2}{\Delta}}
{\lkV{\Gamma}{}{(V_1,V_2):P_1\otimes P_2}{\Delta}}\quad\quad
\seq{\lkV{\Gamma}{}{V_1:P_1}{\Delta}}{\lkV{\Gamma}{}{\inl{V_1}:P_1\oplus P_2}{\Delta}}\quad\quad
\seq{\lkV{\Gamma}{}{V_2:P_2}{\Delta}}{\lkV{\Gamma}{}{\inr{V_2}:P_1\oplus P_2}{\Delta}}$}

\medskip
\fbox{$\seq{c:(\Gamma\vdash \alpha:P,\Delta)}{\lke{\Gamma}{\tilde\mu\alpha^\bullet.c:\notP{P}}{}{\Delta}}\quad\quad
\seq{c:(\Gamma,x_1:P_1,x_2:P_2\vdash\Delta)}{\lke{\Gamma}{\tilde\mu(x_1,x_2).c:P_1\otimes P_2}{}{\Delta}}\quad\quad
\seq{c_1:(\Gamma,x_1:P_1\vdash\Delta)\quad\quad c_2:(\Gamma,x_2:P_2\vdash\Delta)}
{\lke{\Gamma}{\tilde\mu\copaireP{\inl{x_1}.c_1|\inr{x_2}.c_2}:P_1\oplus P_2}{}{\Delta}}$}

\medskip
\fbox{$\seq{\ldots\quad \lkV{\Gamma}{}{V:P}{\Delta}\quad\ldots\quad
\lke{\Gamma}{e:Q}{}{\Delta}\quad\ldots\quad
c:(\lkc{\Gamma\,\ldots,q:P,\ldots}{}{}{\Delta,\ldots,\alpha:Q,\ldots})}
{\Subm{c}{\ldots,\Subi{q}{V},\ldots,\Subi{\alpha}{e}}:(\lkc{\Gamma}{}{}{\Delta})}\quad\quad(\mbox{idem }\Subm{v}{\sigma},\Subm{V}{\sigma},\Subm{e}{\sigma})$}
\end{center}
\end{figure}
\begin{proposition} \label{LKQ-complete}
 If $\Gamma\vdash\Delta$ is provable in $\LK$, then it is provable in $\LKQ$.
\end{proposition}

\Proof Since we have defined a syntax for $\LK$ proofs in section \ref{LK-proofs-sec}, all we have to do is to translate this syntax into the focalised one.  All cases are obvious (only inserting the coercion from values to expressions where appropriate) except for the introduction of $\otimes$ and $\oplus$ on the right, for which we can define $\inl{\mu\alpha_1.c_1}$
as
\begin{center}
$
\lkv{\Gamma}{}{\mu\alpha.\coupe{\mu\alpha_1.c_1}{\tilde\mu x_1.\coupe{\Vtov{(\inl{x_1})}}
{\alpha}}:P_1\oplus P_2}{\Delta}  \quad\quad(\mbox{idem {\it inr}})$
\end{center}
and $(\mu\alpha_1.c_1,\mu\alpha_2.c_2)$ as
$(\lkv{\Gamma}{}{\mu\alpha.\coupe{\mu\alpha_2.c_2}
{\tilde\mu x_2.\coupe{\mu\alpha_1.c_1}
{\tilde\mu x_1.\coupe{\Vtov{(x_1,x_2)}}
{\alpha}}}:P_1\otimes P_2}{\Delta})$. 
\qed

\medskip
We make two observations on the translation involved in the proof of Proposition \ref{LKQ-complete}.
\smallskip\noindent
\begin{remark} \label{LK-LKQ-choice}
The translation {\em introduces cuts}: in particular, a cut-free proof is translated to a proof with (lots of) cuts. It also {\em fixes an order of evaluation}: one should read the translation of right introduction as a protocol prescribing the evaluation of the second element of a pair and then of the first (the pair is thus in particular {\em strict}, as observed in \cite{Munch2009}) (see also
  \cite{ZeilbergerCU,Levy2004}). An equally reasonable choice would have been to permute the two $\tilde\mu$s: that would have encoded a left-to-right order of evaluation. This non-determinism of the translation has been known ever since   Girard's seminal work  \cite{GirardLC}.
\end{remark}
\begin{remark} \label{LK-LKQ-not-red-refl}
The translation is not reduction-preserving, which is expected (since focalisation induces
restrictions on the possible reductions), but it is not reduction-reflecting either, in the sense that new reductions are possible on the translated terms. Here is an example (where, say $\mu\_.c$ indicates a binding with a dummy (i.e., fresh) variable). The translation of
$\coupe{(\mu\_.c_1,\mu\_.c_2)}{\tilde\mu x.c_3}$ rewrites to (the translation of) $c_2$:
\begin{center}
$\coupe{\mu\alpha.\coupe{\mu\_.c_2}
{\tilde\mu x_2.\coupe{\mu\_.c_1}
{\tilde\mu x_1.\coupe{\Vtov{(x_1,x_2)}}
{\alpha}}}}{\tilde\mu x.c_3}\quad\longrightarrow^*\quad
\coupe{\mu\alpha.c_2}{\tilde\mu x.c_3}\quad\longrightarrow^*\quad c_2
$
\end{center}
while the source term is blocked.
If we wanted to cure this, we could turn Proposition \ref{LKQ-complete}'s encodings into additional rewriting rules in the source language. We refrain to do so, since we were merely interested in the source syntax as a stepping stone for the focalised one, and we are content that on one hand the rewriting system of Section \ref{LK-proofs-sec} was good enough to eliminate cuts, and that on the other hand the focalised system is complete with respect to provability.  But we note that the same additional rules {\em do} appear in the {\em target} language  (and are called $\varsigma$-rules, after \cite{Wadlerdual}) in \cite{Munch2009}. This is because in  the focalised syntax proposed in \cite{Munch2009} there is no restriction on the terms of the language, hence $(\mu\_.c_1,\mu\_.c_2)$ is a legal term.
\end{remark}

\smallskip
We move on to cut elimination, which (cf. Section \ref{LK-proofs-sec}) is expressed by means of three sets of rewriting rules, given in Figure 2.
Note that we now have only one way to reduce $\coupe{\mu\alpha.c_1}{\tilde\mu x.c_2}$ (no more critical pair).
As already stressed in Section \ref{LK-proofs-sec}), the commutation rules
 are the usual rules defining (capture-avoiding) substitution.
 The overall operational semantics   features call-by-value by the fact that variables $x$ receive values, 
 and  features also call-by-name (through symmetry, see the logic $\mathsf{LKT}$ in Section \ref{encodings-sec}) by the fact that continuation variables $\alpha$ receive contexts.
 
 \begin{figure}[t] \label{LKQ-cut-elim-rules}
\caption{Cut eliminition in $\mathsf{L}_{\textrm{foc}}$}
$$\begin{array}{lllll}
(\mbox{control}) && \coupe{\mu\alpha.c}{e}  \longrightarrow  \Sub{c}{\alpha}{e}
\quad\quad\quad\quad\;\; \coupe{\Vtov{V}}{\tilde\mu x.c}  \longrightarrow  \Sub{c}{x}{V}\\

(\mbox{logical}) &&\coupe{\Vtov{(e^\bullet)}}{\tilde\mu \alpha^\bullet.c}  \longrightarrow  \Sub{c}{\alpha}{e}
\quad\quad\quad\quad\quad\quad\quad\quad\quad\;\;
\quad\quad\quad \coupe{\Vtov{(V_1,V_2)}}{\tilde\mu(x_1,x_2).c} \longrightarrow  \Subm{c}{\Subi{x_1}{V_1},\Subi{x_2}{V_2}}\\
&& \coupe{\Vtov{\inl{V_1}}}{\tilde\mu\copaireP{\inl{x_1}.c_1|\inr{x_2}.c_2}}  \longrightarrow  \Sub{c_1}{x_1}{V_1}
\quad\quad \coupe{\Vtov{\inr{V_2}}}{\tilde\mu\copaireP{\inl{x_1}.c_1|\inr{x_2}.c_2}}  \longrightarrow  \Sub{c_2}{x_2}{V_2}\\
(\mbox{commutation}) &&
\Subm{\coupe{v}{e}}{\sigma}\longrightarrow \coupe{\Subm{v}{\sigma}}{\Subm{e}{\sigma}} \quad\mbox{etc}\ldots\\
\end{array}$$
\Figbar
\end{figure}

 The reduction system presented in Figure 2 is {\em confluent}, as it is an orthogonal system in the sense of higher-order rewriting systems (left-linear rules, no critical pairs) \cite{NipkHORS}.
\begin{remark} \label{mu-cosmetic}
{\it About $\mu$}: we note that an expression  $\mu\beta.c$ is used only in a command $\coupe{\mu\beta.c}{e}$, and in such a context it can be expressed as
$\coupe{\Vtov{(e^\bullet)}}{\tilde\mu\beta^\bullet.c}$, which indeed reduces to $\Sub{c}{\beta}{e}$. However, using such an encoding would mean to shift from a direct to an indirect style for terms of the form $\mu\alpha.c$.
\end{remark}
\begin{proposition} \label{LKQ-cut-elim-prop}
Cut-elimination holds in $\LKQ$.
\end{proposition}
\Proof This is an easy consequence of the following three properties:

\smallskip\noindent
1) {\it Subject reduction.} This is checked as usual rule by rule.

\smallskip\noindent
2) {\it Weak normalisation.} One first gets rid of the redexes $\coupe{\mu\alpha.c}{e}$ by reducing them all  (no such redex is ever created by the other reduction rules). 
As usual, one measures cuts by the size of the cut formula, called the degree of the redex, and at each step of normalisation, one chooses a redex of maximal degree all of whose subredexes have striclty lower degree.   We then package reductions by considering $\coupe{\Vtov{V}}{\tilde\mu x.c}\longrightarrow \Subwn{c}{\Subi{x}{V}}$
 (idem for the logical rules) as a single step, where $\Subwn{c}{\sigma}$ is an augmented (implicit) substitution, defined by induction as usually except for $\coupe{v}{e}$:
$$\begin{array}{l}
\Subwn{\coupe{\Vtov{x}}{\tilde\mu\alpha^\bullet.c}}{\Subi{x}{e^\bullet},\sigma} = \Subwn{c}{\Subi{\alpha}{e},\Subi{x}{e^\bullet},\sigma}\\
\Subwn{\coupe{\Vtov{x}}{\tilde\mu (x_1,x_2).c}}{\Subi{x}{(V_1,V_2)},\sigma}= 
\Subwn{c}{\Subi{x_1}{V_1},\Subi{x_2}{V_2},\Subi{x}{(V_1,V_2)},\sigma}\\
\Subwn{\coupe{\Vtov{x}}{\tilde\mu[\inl{x_1}.c_1|\inr{x_2}.c_2]}}{\Subi{x}{\inl{V_1}},\sigma}= 
\Subwn{c}{\Subi{x_1}{V_1},\Subi{x}{\inl{V_1}},\sigma} \quad(\mbox{idem {\it inr}})\\
\Subwn{\coupe{v}{e}}{\sigma} = \coupe{\Subwn{v}{\sigma}}{\Subwn{e}{\sigma}} \quad\mbox{otherwise}
\end{array}$$
This is clearly a well-founded definition, by induction on the term in which substitution is performed
(whatever the substitution is).
This new notion of reduction ensures the following property:  is $t_1 \longrightarrow t_2$ is obtained by reducing $R_1$ in $t_1$ and if $R_2$ is a redex created in $t_2$ by this (packaged) reduction, then $R_1$ is of the form $\coupe{e^\bullet}{\tilde\mu\alpha^\bullet.c}$, where $c$ contains a subterm   $\coupe{\Vtov{V}}{\alpha}$, which becomes $R_2$ in $t_2$.  The key property is then that the degree of the created redex  (the size of some formula $P$) is strictly smaller than the degree of the creating one (the size of $\notP{P}$)\footnote{If we had not packaged reduction, we would have had to deal with
the creation of  redexes, say by susbstitution of some $V$ for $x$, where the substitution could have been launched by firing a redex of the same degree as the created one.}.
The other useful property is that residuals of redexes preserve their degree.  Then the argument is easily concluded by associating to each term as global measure the multiset of the degrees of its redexes. This measure strictly decreases at each step (for the multiset extension of the ordering on natural numbers).

\smallskip\noindent
3) {\it Characterisation of normal forms.}  A command  in normal  form  has one of the following shapes (all contractions):
$$\coupe{\Vtov{V}}{\alpha} \quad\quad\quad \coupe{\Vtov{x}}{\tilde\mu\alpha^\bullet.c}\quad\quad
\coupe{\Vtov{x}}{\tilde\mu(x_1,x_2).c}\quad\quad
\coupe{\Vtov{x}}{\tilde\mu[\inl{x_1}.c_1|\inr{x_2}.c_2]}  \quad\quad\quad\quad\qed$$

\begin{corollary} Every sequent $\Gamma\vdash\Delta$ that is provable in $\LK$ admits a (cut-free) proof respecting the focalised discipline.
\end{corollary}
\Proof  Let $\pi$ be a proof of $\Gamma\vdash\Delta$. By Proposition
\ref{LKQ-complete}, $\pi$ translates to a command $c:(\lkc{\Gamma}{}{}{\Delta})$, which by Proposition \ref{LKQ-cut-elim-prop} reduces to a term denoting a cut-free proof. The $\LK$ proof obtained by erasing meets the requirement\footnote{This argument of focalisation via normalisation goes back to \cite{GirardLC} (see also \cite{LaurentFoc} for a detailed proof in the case of linear logic).}. 
\qed

\smallskip
Also, by confluence and weak normalisation, $\LKQ$  is computationally coherent:
$(\lkV{x:P,y:P}{}{x:P}{})$ and $(\lkV{x:P,y:P}{}{y:P}{})$ are not provably equal, being normal forms.

\smallskip
Our syntactic choices in this paper have been guided by the phases of focalisation. Indeed, with our syntax, the focalised proof search cycle can be represented as follows (following a branch from the root):
{\small $$\begin{array}{lllllll}
(\mbox{right phase}) &&
\coupe{\Vtov{V}}{\alpha}:(\lkc{\Gamma}{}{}{\alpha:P,\Delta}) \;\rightsquigarrow_{-/+} \;\lkV{\Gamma}{}{V:P}{\alpha:P,\Delta}\;
\rightsquigarrow_{+}^*\; \lkV{\Gamma}{}{(\tilde\mu x.c)^\bullet:\notP{Q}}{\Delta}\\
&&\quad \rightsquigarrow_{+/-}\; \lke{\Gamma}{\tilde\mu x.c:Q}{}{\Delta}\;
\rightsquigarrow_{-} \; c:(\lkc{\Gamma,x:Q}{}{}{\Delta})   \quad\quad(\mbox{idem other $\tilde\mu$ binders})\\
(\mbox{left phase}) && \coupe{\Vtov{x}}{\tilde\mu\alpha^\bullet.c}:(\lkc{\Gamma,x:\notP{P}}{}{}{\Delta})
 \; \rightsquigarrow_{-}^* \; c:(\lkc{\Gamma,x:\notP{P}}{}{}{\alpha:P,\Delta})
\\
&&
\coupe{\Vtov{x}}{\tilde\mu(x_1,x_2).c}:(\lkc{\Gamma,x:P_1\otimes P_2}{}{}{\Delta})
 \; \rightsquigarrow_{-}^* \; c:(\lkc{\Gamma,x_1:P_1,x_2:P_2,x:P_1\otimes P_2}{}{}{\Delta})\\
&&
\coupe{\Vtov{x}}{\tilde\mu[\inl{x_1}.c_1|\inr{x_2}.c_2]}:(\lkc{\Gamma,x:P_1\oplus P_2}{}{}{\Delta})
 \; \rightsquigarrow_{-}^* \; c_1:(\lkc{\Gamma,x_1:P_1,x:P_1\oplus P_2}{}{}{\Delta})\\
 &&
\coupe{\Vtov{x}}{\tilde\mu[\inl{x_1}.c_1|\inr{x_2}.c_2]}:(\lkc{\Gamma,x:P_1\oplus P_2}{}{}{\Delta})
 \; \rightsquigarrow_{-}^* \; c_2:(\lkc{\Gamma,x_2:P_2,x:P_1\oplus P_2}{}{}{\Delta})
\end{array}
$$}
Note that values and commands correspond to positive and negative phases, respectively. The other two categories of terms act as intermediates. 

\smallskip
We can also add $\eta$-equivalences (or expansion rules, when read from right to left)  to the system, as follows (where all mentioned variables are fresh for the mentioned terms):
\begin{center}
$\begin{array}{lll}
\mu\alpha.\coupe{v}{\alpha}=v &\quad\quad\quad &  \tilde\mu(x_1,x_2).\coupe{\Vtov{(x_1,x_2)}}{e}=e\\
 \tilde\mu x.\coupe{\Vtov{x}}{e}=e &\quad\quad\quad & \tilde\mu[\inl{x_1}.\coupe{\Vtov{\inl{x_1}}}{e}|\inr{x_2}.\coupe{\Vtov{\inr{x_2}}}{e}]=e\\
&\quad\quad\quad& \tilde\mu\alpha^\bullet\coupe{\Vtov{(\alpha^\bullet)}}{e} =e
\end{array}$
\end{center}
The rules on the left column allow us to cancel a deactivation followed by an activation (the control rules do the job for the sequence in the reverse order), while the rules in the right column express the reversibility of the negative rules.

\begin{example} \label{LKQ-exp-ex}
We relate
$(\notP{P_1})\otimes(\notP{P_2})$ and $\notP{(P_1\oplus P_2)}$ 
(cf. the well-know isomorphism of linear logic, reading $\notP{P}$ as $!\overline{P}$).
There exist 
$$\begin{array}{l}
c_1:(\lkc{y:\notP{(P_1\oplus P_2)}}{}{}{\alpha:\notP{P_1}\otimes\notP{P_2}})\quad\quad
c_2:(\lkc{x:\notP{P_1}\otimes\notP{P_2}}{}{}{\gamma:\notP{(P_1\oplus P_2)}})
\end{array}$$
such that, say\footnote{In the $\lambda$-calculus a provable isomorphism  is a pair $(x:A\vdash v:B)$, $(y:B\vdash w:A)$ such that $\Subimpl{w}{y}{v}$ reduces to $x$ (and conversely). Here, we express this substitution (where $v,w$ are not values) as
$\mu\alpha.\coupe{v}{\tilde\mu y.\coupe{w}{\alpha}}$, and the reduction as
$\coupe{v}{\tilde\mu y.\coupe{w}{\alpha}}\longrightarrow^*\coupe{\Vtov{x}}{\alpha}$.
}  $\coupe{\mu\gamma.c_2}{\tilde\mu y.c_1}$, reduces
 $\coupe{\Vtov{x}}{\alpha}:(\lkc{x:\notP{P_1}\otimes\notP{P_2}}{}{}{\alpha:\notP{P_1}\otimes\notP{P_2}})$ . We set
$$\begin{array}{lll}
V_1=((\tilde\mu y'_1.\coupe{\Vtov{\inl{y'_1}}}{\beta})^\bullet,(\tilde\mu y'_2.\coupe{\Vtov{\inr{y'_2}}}{\beta})^\bullet)  &\quad\quad\quad& \lkV{}{}{V_1:\notP{P_1}\otimes\notP{P_2}}{\beta:P_1\oplus P_2}\\
V_2=(\tilde\mu\copaireP{\inl{y_1}.\coupe{\Vtov{y_1}}{\alpha_1}|\inr{y_2}.\coupe{\Vtov{y_2}}{\alpha_2}})^\bullet &\quad\quad&
 \lkV{}{}{V_2:\notP{(P_1\oplus P_2)}}{\alpha_1:P_1,\alpha_2:P_2}
\end{array}$$
We take $c_1=\coupe{\Vtov{y}}{\tilde\mu\beta^\bullet.
\coupe{\Vtov{V_1}}{\alpha}}$ and 
$c_2=\coupe{\Vtov{x}}{\tilde\mu(\alpha_1^\bullet,\alpha_2^\bullet).
\coupe{\Vtov{V_2}}{\gamma}}$,
where  $\tilde\mu(\alpha_1^\bullet,\alpha_2^\bullet).c$ is defined as a shorthand for, say
$\tilde\mu(x_1,x_2).\coupe{\Vtov{x_2}}{\tilde\mu\alpha_2^\bullet.\coupe{\Vtov{x_1}}{\tilde\mu\alpha_1^\bullet .c}}$.
We have:
$$\begin{array}{lll}
\coupe{\mu\gamma.c_2}{\tilde\mu y.c_1} & \longrightarrow^* &  \coupe{\Vtov{x}}{\tilde\mu(\alpha_1^\bullet,\alpha_2^\bullet).
\coupe{((\tilde\mu y'_1.\coupe{\Vtov{(y')_1}}{\alpha_1})^\bullet,(\tilde\mu y'_2.\coupe{\Vtov{(y')_2}}{\alpha_2}^\bullet)}{\alpha}}\\
&& = \coupe{\Vtov{x}}{\tilde\mu(\alpha_1^\bullet,\alpha_2^\bullet).
\coupe{(\alpha_1^\bullet,\alpha_2^\bullet)}{\alpha}}\\
&& =  \coupe{\Vtov{x}}{\alpha}
\end{array}$$
\end{example}

We end the section with a lemma that will be useful in Section \ref{LKQS-sec}.
\begin{lemma}  \label{e-subst-lemma}
\begin{itemize} \item If $\lke{\Gamma,x:\notP{P}}{e:Q}{}{\Delta}$, then
$\lke{\Gamma}{\Subimpl{e}{x}{\alpha^\bullet}:Q}{}{\alpha:P,\Delta}$.
\item If $\lke{\Gamma,x:P_1\otimes P_2}{e:Q}{}{\Delta}$, then
$\lke{\Gamma,x_1:P_1,x_2:P_2}{\Subimpl{e}{x}{(x_1,x_2)}:Q}{}{\Delta}$.
\item If $\lke{\Gamma,x:P_1\oplus P_2}{e:Q}{}{\Delta}$, then
$\lke{\Gamma,x_1:P_1}{\Subimpl{e}{x}{\inl{x_1}}:Q}{}{\Delta}$ and 
$\lke{\Gamma,x_2:P_2}{\Subimpl{e}{x}{\inr{x_2}}:Q}{}{\Delta}$.
\end{itemize}
(and similarly for $c,V,v$),
where $\Subimpl{t}{x}{V}$ (resp. $\Subimpl{t}{\alpha}{e}$) denotes the usual substitution (cf. Section \ref{introduction}).
\end{lemma}

\section{Encodings} \label{encodings-sec}

\noindent{\bf Encoding CBV $\lbd$($\mu$)-calculus into $\mathsf{LKQ}$.}
We are now in a position to hook up with the material of Section \ref{intro-sec}. We
 can encode the call-by-value $\lbd$-calculus, by defining the following derived CBV implication  and terms:
$$\begin{array}{c}
P\rightarrow^v Q=\notP{(P\:\otimes\:\notP{Q})}\\
\lbd x.v=\Vtov{((\tilde\mu(x,\alpha^\bullet).\coupe{v}{\alpha})^\bullet)}\quad\quad\quad
v_1v_2=\mu\alpha.\coupe{v_2}{\tilde\mu x.\coupe{v_1}{\Vtoe{(\Vtov{(x,\alpha^\bullet)})}}}
\end{array}$$
where $\tilde\mu(x,\alpha^\bullet).c$ is the abbreviation used in Example \ref{LKQ-proofs-ex} and where $\Vtoe{V}$ stands for $\tilde\mu\alpha^\bullet.\coupe{\Vtov{V}}{\alpha}$.
These definitions provide us with a translation,
which extends to (call-by-value) $\lambda\mu$-calculus \cite{OngStew,Rocheteau}, and factors though $\lbdqch$-calculus (cf. Section \ref{intro-sec}), defining $V\cdot e$ as
$\Vtoe{(V,e^\bullet)}$\footnote{In \cite{CH2000} we also had a difference operator $B-A$ (dual to implication), and two associated introduction operations, whose encodings in the present syntax are 
$\beta\lambda.e=\tilde\mu(\beta^\bullet,x).\coupe{\Vtov{x}}{e}$  and $e\cdot V=(e^\bullet,V)$.}.
The translation makes also sense in the untyped setting, as the following example shows.
\begin{example} \label{Delta-Delta}
Let $\Delta=\lbd x.xx$. We have $\CBVP{\Delta\Delta}=\mu\gamma.c$,  and $c\longrightarrow^* c$, with
 $$c=\coupe{\Vtov{(e^\bullet)}}{\tilde\mu z.\coupe{\Vtov{(e^\bullet)}}{\Vtoe{(z,\gamma^\bullet)}}}\quad\mbox{ and }\quad e=\tilde\mu(x,\alpha^\bullet).\coupe{\Vtov{x}}{\tilde\mu y.\coupe{\Vtov{x}}{\Vtoe{(y,\alpha^\bullet)}}}$$
\end{example}

\noindent
{\bf Encoding CBN $\lbd$($\mu$)-calculus.}
What about CBN?  We can translate it to $\mathsf{LKQ}$, but at the price of translating terms to contexts, which is a violence to our goal of giving an  intuitive semantics to the first abstract machine presented in Section \ref{intro-sec}.  Instead, we spell out the system dual to $\mathsf{LKQ}$, which is known as $\mathsf{LKT}$, in which expressions and contexts will have negative types, and in which we shall be able to express CBN $\lbd$-terms as expressions.  Our syntax for 
$\mathsf{LKT}$ is a mirror image of that for $\mathsf{LKQ}$: it exchanges the $\mu$ and $\tilde\mu$, the $x$'s and the $\alpha$'s, etc..., and renames {\it inl}, {\it inr} as {\it fst}, {\it snd}, which are naturally associated with $\with$ while the latter were naturally associated with $\oplus$:
$$\begin{array}{lll}
\mbox{Commands} && c::=\coupe{v}{e}\\
\mbox{Covalues} && E::= \alpha \Alt  [E,E] \Alt \fst{E}\Alt \snd{E}\Alt v^\bullet\\
\mbox{Contexts} && e::= \Etoe{E} \Alt \tilde\mu x.c\\
\mbox{Expressions} && v::=x \Alt \mu \alpha.c\Alt \mu x^\bullet.c \Alt\ldots 
\end{array}$$
Note that focalisation is now on the left, giving rise to a syntactic category of {\em covalues} (that
were called applicative contexts in \cite{CH2000}).\footnote{Note also that the duality sends a  command $\coupe{v}{e}$ to a command $\coupe{e'}{v'}$ where $v'$, $e'$ are the mirror images of $v$, $e$.}

The rules are all obtained from $\mathsf{LKQ}$ by duality:
$$\begin{array}{c}
\seq{}{\lkE{\Gamma}{\alpha:N}{}{\Delta\:,\:\alpha:N}}\quad\quad
\seq{\lkE{\Gamma}{E_1:N_1}{}{\Delta}}{\lkE{\Gamma}{\fst{E_1}:N_1\with N_2}{}{\Delta}}
\\\\
\seq{}{\lkv{\Gamma\:,\:x:N}{}{x:N}{\Delta}}\quad\quad\seq{\lkv{\Gamma}{}{v:N}{\Delta}\quad\quad
\lke{\Gamma}{e:N}{}{\Delta}}
{\coupe{v}{e}:(\lkc{\Gamma}{}{}{\Delta})}   \quad\quad\quad  \ldots 
\end{array}$$

We would have arrived to this logic naturally if we had chosen in Section \ref{LK-proofs-sec} to present $\mathsf{LK}$ with a reversible disjunction on the right and an irreversible conjunction on the left, and in Section \ref{LKQ-sec} to present a focalisation discipline with 
focusing on formulas on the left. 

In $\mathsf{LKT}$ we can define the following derived CBN implication  and terms:
$$\begin{array}{c}
M\rightarrow^n N=(\notN{M})\:\lpar\:N\\
\lbd x.v=\mu( x^\bullet,\alpha).\coupe{v}{\Vtov{\alpha}}\quad\quad\quad
v_1v_2=\mu\alpha.\coupe{v_1}{\Vtov{(v_2^\bullet,\alpha)}}
\end{array}$$
The translation extends to $\lambda\mu$-calculus \cite{Parigot92} and factors though the $\lbdtch$-calculus of \cite{CH2000}, defining $v \cdot E$ as $(v^\bullet,E)$. Note that the covalues involved in executing call-by-name $\lambda$-calculus are just {\em stacks} of expressions (cf. Section \ref{intro-sec}).

With these definitions, we have:
$$\begin{array}{lll}
 \coupe{\lbd x.v_1}{\Vtov{(v_2\cdot E)}} =
 \coupe{\mu( x^\bullet,\alpha).\coupe{v_1}{\Vtov{\alpha}}}{\Vtov{(v_2^\bullet,E)}}
 \longrightarrow  \coupe{\Sub{v_1}{x}{v_2}}{\Vtov{E}}\\
 \coupe{v_1v_2}{\Vtov{E}} = \coupe{\mu\alpha.\coupe{v_1}{\Vtov{(v_2^\bullet,\alpha)}}}{\Vtov{E}}  \longrightarrow \coupe{v_1}{\Vtov{(v_2^\bullet,E)}} = 
 \coupe{v_1}{\Vtov{(v_2\cdot E)}} 
\end{array}$$
We are thus back on our feet (cf. section 1)!

\medskip\noindent
{\bf Translating $\mathsf{LKQ}$ into $\mathsf{NJ}$.}
 Figure 3
presents a translation from $\LKQ$ to intuitionistic natural deduction $\mathsf{NJ}$, or, via Curry-Howard, to $\lbd$-calculus extended with products and sums.
In the translation, $R$ is a fixed target formula (cf. Section \ref{intro-sec}).  We  translate $(\notP{\_})$ as ``$\_$ implies $R$''  (cf. \cite{Krivine91,LRS93}).
We write $B^A$ for function types / intuitionistic implications.
The rules of $\mathsf{L}_{\textrm{foc}}$ are simulated by $\beta$-reductions. One may 
think of the source  $\mathsf{L}_{\textrm{foc}}$ terms as a description of  the target ones ``in  direct style'' (cf. \cite{DanvyBack}).

\begin{figure}[t] \label{LKQ-to-NJ}
\caption{Translation of $\LKQ$ into the $\lbd$-calculus / $\mathsf{NJ}$}

\smallskip

Translation of formulas:
$$\begin{array}{ll}
\CPSP{X}=X & \CPSP{(\notP{P})}=R^{\CPSP{P}}\\
\CPSP{(P\otimes Q)}=(\CPSP{P})\times (\CPSP{Q}) &
\CPSP{P\oplus Q}=(\CPSP{P})+(\CPSP{Q})
\end{array}$$

\smallskip
\noindent
Translation of terms:
$$\begin{array}{l}
\CPSP{{\coupe{v}{e}}}=(\CPSP{v})(\CPSP{e})\quad\quad
\CPSP{(\Vtov{V})}=\lbd k.k(\CPSP{V})  \quad\quad
 \CPSP{(\mu\alpha.c)}=\lambda\contvartovar{\alpha}.(\CPSP{c})=\CPSP{(\tilde\mu\alpha^\bullet.c)}\\
 \CPSP{x}=x \quad\quad \CPSP{(V_1,V_2)}=(\CPSP{(V_1)},\CPSP{(V_2)})\quad\quad
 \CPSP{\inl{V_1}}=\inl{\CPSP{(V_1)}}\quad\quad
  \CPSP{\inr{V_2}}=\inr{\CPSP{(V_2)}}\quad\quad\CPSP{(e^\bullet)}=\CPSP{e}\\
\CPSP{\alpha}=\contvartovar{\alpha}\quad\quad\CPSP{(\tilde\mu x.c)}=\lbd x.(\CPSP{c})\quad\quad
\CPSP{(\tilde\mu(x_1,x_2).c)}=\lbd (x_1,x_2).(\CPSP{c})\\
\CPSP{( \tilde\mu\copaireP{\inl{x_1}.c_1|\inr{x_2}.c_2})}=
\lbd z.\fcase{z}{\fcasei{\inl{x_1}}{\CPSP{(c_1)},\fcasei{\inr{x_2}}{\CPSP{(c_2)}}}}
\end{array}$$

\Figbar
\end{figure}
\begin{proposition} We set
$\CPSP{\Gamma}=\setc{x:\CPSP{P}}{x:P\in\Gamma}\quad\quad R^{\CPSP{\Delta}}=\setc{\contvartovar{\alpha}:R^{\CPSP{P}}}{\alpha:P\in\Delta}$.
We have:
$$\begin{array}{|cc|cc|cc|c|}
\hline
c:(\lkc{\Gamma}{}{}{\Delta}) &&
\lkV{\Gamma}{}{V:P}{\Delta} &&
\lkv{\Gamma}{}{v:P}{\Delta} &&
\lke{\Gamma}{e:P}{}{\Delta}\\
\Downarrow &&\Downarrow &&\Downarrow &&\Downarrow \\
\CPSP{\Gamma}\:,\:R^{\CPSP{\Delta}}\vdash \CPSP{c}:R  &&
\CPSP{\Gamma}\:,\:R^{\CPSP{\Delta}}\vdash \CPSP{V}:\CPSP{P} &&
\CPSP{\Gamma}\:,\:R^{\CPSP{\Delta}}\vdash \CPSP{v}:R^{R^{\CPSP{P}}} &&
\CPSP{\Gamma}\:,\:R^{\CPSP{\Delta}}\vdash \CPSP{e}:R^{\CPSP{P}}\\
\hline
\end{array}$$

Moreover, the translation preserves  reduction: if $t\longrightarrow t'$, then
$\CPSP{t}\longrightarrow^*\CPSP{(t')}$.
\end{proposition}
Composing the  previous translations, from CBN $\lambda\mu$-calculus to $\mathsf{LKT}$ then through duality to $\mathsf{LKQ}$ then to $\mathsf{NJ}$, what we obtain is the CPS translation due to 
Lafont, Reus, and Streicher \cite{LRS93} (LRS translation, for short).


\medskip\noindent
{\bf Translating $\mathsf{LKQ}$ into $\mathsf{LLP}$.}
The translation just given from $\mathsf{LKQ}$ to  $\mathsf{NJ}$ does actually two transformations for the price of one: {\em from classical to intuitionistic}, and {\em from sequent calculus style to natural deduction style}.  The intermediate target and source of this decomposition is nothing but a subsystem of  Laurent's polarised linear logic $\mathsf{LLP}$ \cite{LaurentTh}
\footnote{Specifically,  no positive formula is allowed on the right in  the rules for $\lpar$ and $\with$ and in the right premise of the cut rule.}. 
We adopt a presentation of  $\mathsf{LLP}$ in which all negative formulas are handled as positive formulas on the left, and hence in which $!N$ and $?P$ are replaced by $\notP{}$ with the appropriate change of side. With these conventions, $\mathsf{LLP}$ is nothing but the system called $\mathsf{LJ}_0$ in \cite{Laurent2009}. 
For the purpose of giving a system $\mathsf{L}$ term syntax, we distinguish  three kinds of sequents for our subsystem of $\mathsf{LLP}$:
$$(\lkc{\Gamma}{}{}{})\quad\quad(\lkV{\Gamma}{}{P}{})\quad\quad
(\lke{\Gamma}{P}{}{})$$
The syntax, the computation rules, and the typing rules are as follows (omitting explicit substitutions):
$$\begin{array}{l}
c::=\coupe{V}{e}\quad V::= x\Alt e^\bullet \Alt (V,V) \Alt \inl{V} \Alt \inr{V}\\ e::=\derel{V}\Alt \tilde\mu x.c \Alt \tilde\mu(x_1,x_2).c\Alt \tilde\mu[\inl{x_1}.c_1|\inr{c_2}.c_2]
\end{array}
$$

$$\begin{array}{lllll}
\coupe{V}{\tilde\mu x.c}  \longrightarrow  \Sub{c}{x}{V}\\
\coupe{e^\bullet}{\derel{V}}  \longrightarrow  \coupe{V}{e} \\
\coupe{(V_1,V_2)}{\tilde\mu(x_1,x_2).c} \longrightarrow  \Subm{c}{\Subi{x_1}{V_1},\Subi{x_2}{V_2}}\\
 \coupe{\inl{V_1}}{\tilde\mu\copaireP{\inl{x_1}.c_1|\inr{x_2}.c_2}}  \longrightarrow  \Sub{c_1}{x_1}{V_1}
\quad\quad \coupe{\inr{V_2}}{\tilde\mu\copaireP{\inl{x_1}.c_1|\inr{x_2}.c_2}}  \longrightarrow  \Sub{c_2}{x_2}{V_2}
\end{array}$$

$$\seq{}{\lkV{\Gamma\:,\:x:P}{}{x:P}{}}\quad\quad
\seq{\lkV{\Gamma}{}{V:P}{}\quad\quad
\lke{\Gamma}{e:P}{}{}}
{\coupe{V}{e}:(\lkc{\Gamma}{}{}{})}\quad\quad
\seq{c:(\lkc{\Gamma\:,\:x:P}{}{}{})}{\lke{\Gamma}{\tilde\mu x.c:P}{}{}}
$$
$$\seq{\lke{\Gamma}{e:P}{}{}}{\lkV{\Gamma}{}{e^\bullet:\notP{P}}{}}\quad\quad\seq{\lkV{\Gamma}{}{V_1:P_1}{} \quad\quad \lkV{\Gamma}{}{V_2:P_2}{}}
{\lkV{\Gamma}{}{(V_1,V_2):P_1\otimes P_2}{}}\quad\quad
\seq{\lkV{\Gamma}{}{V_1:P_1}{}}{\lkV{\Gamma}{}{\inl{V_1}:P_1\oplus P_2}{}}\quad\quad
\seq{\lkV{\Gamma}{}{V_2:P_2}{}}{\lkV{\Gamma}{}{\inr{V_2}:P_1\oplus P_2}{}}$$
$$\quad\seq{\lkV{\Gamma}{}{V:P}{}}{\lke{\Gamma}{\derel{V}:\notP{P}}{}{}}\quad\quad
\seq{c:(\Gamma,x_1:P_1,x_2:P_2\vdash)}{\lke{\Gamma}{\tilde\mu(x_1,x_2).c:P_1\otimes P_2}{}{}}\quad\quad
\seq{c_1:(\Gamma,x_1:P_1\vdash)\quad\quad c_2:(\Gamma,x_2:P_2\vdash)}
{\lke{\Gamma}{\tilde\mu\copaireP{\inl{x_1}.c_1|\inr{x_2}.c_2}:P_1\oplus P_2}{}{}}$$

\smallskip\noindent

The constructs $e^\bullet$ and $\derel{V}$ transcribe $\mathsf{LLP}$'s 
{\em promotion} and {\em dereliction} rule, respectively.   
The compilation from $\mathsf{LKQ}$ to (the subsystem of) $\mathsf{LLP}$
 turns every $\alpha:P$ on the right to a $\contvartovar{\alpha}:\notP{P}$ on the left. We write
 $\notP{(\ldots,\alpha:P,\ldots)}= (\ldots,k_\alpha:\notP{P},\ldots)$.
 The translation is as follows (we give only the non straightforward cases):
 $$\begin{array}{l}
 \coupe{v}{e}_{\mathsf{LLP}}=\coupe{(e_{\mathsf{LLP}})^\bullet}{v_{\mathsf{LLP}}}\\
 (\mu\alpha.c)_{\mathsf{LLP}}=\tilde\mu\contvartovar{\alpha}.c_{\mathsf{LLP}}\quad\quad
 (\Vtov{V})_{\mathsf{LLP}}= \derel{(V_{\mathsf{LLP}})}\\
 \alpha_{\mathsf{LLP}}=\tilde\mu x.\coupe{\contvartovar{\alpha}}{\derel{x}}\quad
(\tilde\mu\alpha^\bullet.c)_{\mathsf{LLP}}=\tilde\mu\contvartovar{\alpha}.(c_{\mathsf{LLP}}) 
 \end{array}$$
 Note that we can optimise the translation of $\coupe{\Vtov{V}}{e}$, and (up to an expansion rule) of $\Vtoe{V}=\tilde\mu\alpha^\bullet.\coupe{\Vtov{V}}{\alpha}$:
 $$\begin{array}{lll}
 \coupe{\Vtov{V}}{e}_{\mathsf{LLP}}=\coupe{(e_{\mathsf{LLP}})^\bullet}{\derel{(V_{\mathsf{LLP}})}}
 \longrightarrow\coupe{V_{\mathsf{LLP}}}{e_{\mathsf{LLP}}}\\
 (\Vtoe{V})_{\mathsf{LLP}} =
 \tilde\mu \contvartovar{\alpha}.\coupe{V_{\mathsf{LLP}}}{\alpha_{\mathsf{LLP}}}
 \;= \tilde\mu \contvartovar{\alpha}.\coupe{V_{\mathsf{LLP}}}{\tilde\mu x.\coupe{\contvartovar{\alpha}}{\derel{x}}} \longrightarrow  \tilde\mu \contvartovar{\alpha}.\coupe{\contvartovar{\alpha}}{\Vtov{(V_{\mathsf{LLP}})}} = \derel{(V_{\mathsf{LLP}})}
\end{array} $$
These optimisations allow us to define a right inverse to ${}_{\mathsf{LLP}}$ (that maps $\Vtov{V}$ to $\Vtoe{V}$)), i.e.:
\begin{center}
{\em $\mathsf{LLP}$ (restricted as above) appears as a retract of $\mathsf{LKQ}$}.
\end{center}
 The translation simulates reductions and is well typed: 
 $$\begin{array}{lll}
 c:(\lkc{\Gamma}{}{}{\Delta}) & \quad\Rightarrow\quad & c_{\mathsf{LLP}}: (\lkc{\Gamma,\notP{\Delta}}{}{}{})\\
 \lkv{\Gamma}{}{v:P}{\Delta} & \quad\Rightarrow\quad & \lke{\Gamma,\notP{\Delta}}{v_{\mathsf{LLP}}:\notP{P}}{}{}\\
 \lkV{\Gamma}{}{V:P}{\Delta} & \quad\Rightarrow\quad & \lkV{\Gamma,\notP{\Delta}}{}{V_{\mathsf{LLP}}:P}{}\\
\lke{\Gamma}{e:P}{}{\Delta} & \quad\Rightarrow\quad & \lke{\Gamma,\notP{\Delta}}{e_{\mathsf{LLP}}:P}{}{}
 \end{array}$$

We note that this compilation blurs the distinction between  a continuation variable and an ordinary variable (like in the classical CPS translations).

\begin{example}  
The classical proof $(\lke{}{\tilde\mu\beta^\bullet.\coupe{\Vtov{(\alpha^\bullet)}}{\beta}:\notP{\notP{P}}}{}{\alpha:P})$
 is translated (using the above optimisation) to the  intuitionistic proof
 $(\lke{\contvartovar{\alpha}:\notP{P}}{\derel{((\tilde\mu x.\coupe{\contvartovar{\alpha}}{\derel{x}})^\bullet)}:\notP{\notP{P}}}{}{})$. Without term decorations, we have turned a proof of the classically-only provable sequent
 $(\lke{}{\notP{\notP{P}}}{}{P})$ into an intuitionistic proof of $(\lke{\notP{P}}{\notP{\notP{P}}}{}{})$.
\end{example}

\smallskip
All what is left to do in order to reach then $\mathsf{NJ}$\footnote{In fact, the target of the translation uses only implications of the form $R^P$. Seeing $R$ as ``false'', this means that the target is in fact intuitionistic logic with conjunction, disjunction and negation in natural deduction style.}
from  (our subsystem of) $\mathsf{LLP}$ 
is to turn contexts $e:P$ into values of type $\notP{P}$,  and to rename $\notP{\_}$, $\otimes$, and $\oplus$ as $R^{\_},\times$, and $+$, respectively.  More precisely, we describe the target syntax (a subset of a $\lambda$-calculus with sums and products) as follows:
$$\begin{array}{l}
c::= VV\quad\quad V::= x\Alt (V,V) \Alt \inl{V}\Alt\inr{V} \Alt \lbd x.c \Alt \lbd(x_1,x_2).c\Alt \lbd z.\fcase{z}{\fcasei{\inl{x_1}}{c_1,\fcasei{\inr{x_2}}{c_2}}}
\end{array}$$
Again, we give only the non trivial cases of the translation:
$$\begin{array}{l}
\coupe{V}{e}_{\mathsf{NJ}}= (e_{\mathsf{NJ}})(V_{\mathsf{NJ}})\quad\quad (e^\bullet)_{\mathsf{NJ}}=e_{\mathsf{NJ}}\quad\quad
(\derel{V})_{\mathsf{NJ}}=\lambda k.k(V_{\mathsf{NJ}})\\(\tilde\mu x.c)_{\mathsf{NJ}}=\lbd x.(c_{\mathsf{NJ}})\quad (\tilde\mu(x_1,x_2).c)_{\mathsf{NJ}}=\lbd(x_1,x_2).(c_{\mathsf{NJ}})\\
 (\tilde\mu[\inl{x_1}.c_1|\inr{c_2}.c_2])_{\mathsf{NJ}}=\lbd z.\fcase{z}{\fcasei{\inl{x_1}}{((c_1)_{\mathsf{NJ}}),\fcasei{\inr{x_2}}{(c_2)_{\mathsf{NJ}}}}}
\end{array}$$

\begin{proposition}   \label{factor-cps} For all $\mathsf{L}_{\textrm{foc}}$ term  $t$  (where $t::= c\Alt V\Alt v\Alt e$), we have:
$$\CPSP{t}=_{\beta\eta} (t_{\mathsf{LLP}})_{\mathsf{NJ}}\;.$$
\end{proposition}
\Proof  We treat the non trivial cases:
$$\begin{array}{lll}
\coupe{v}{e}_{\mathsf{LLP},\mathsf{NJ}} = \coupe{(e_{\mathsf{LLP}})^\bullet}{v_{\mathsf{LLP}}}_{\mathsf{NJ}} =   (v_{\mathsf{LLP},\mathsf{NJ}})(((e_{\mathsf{LLP}})^\bullet)_{\mathsf{NJ}}) =  (v_{\mathsf{LLP},\mathsf{NJ}})(e_{\mathsf{LLP},\mathsf{NJ}})\\
(\Vtov{V})_{\mathsf{LLP},\mathsf{NJ}} =
( \derel{(V_{\mathsf{LLP}})})_{\mathsf{NJ}} =\lbd k.k(V_{\mathsf{LLP},\mathsf{NJ}} )\\
\alpha_{\mathsf{LLP},\mathsf{NJ}}   =  (\tilde\mu x.\coupe{\contvartovar{\alpha}}{\derel{x}})_{\mathsf{NJ}} = \lbd x.(\coupe{\contvartovar{\alpha}}{\derel{x}}_{\mathsf{NJ}})=\lbd x.((\derel{x})_{\mathsf{NJ}})\contvartovar{\alpha}=\lbd x.(\lbd k.kx)\contvartovar{\alpha}=_\beta\lbd x.\contvartovar{\alpha}x=_\eta \contvartovar{\alpha}\;\qed
\end{array}$$
Note that, in the proof of the above proposition, the $\beta$ step  is a typical ``administrative reduction'', so that morally the statement of the proposition holds with $\eta$ only.

\medskip\noindent
{\bf A short  foray into linear logic.} We end this section by  placing the so-called Girard's 
translation of the call-by-name $\lambda$-calculus to linear logic in perspective. The target of this translation is in fact  the polarised fragment  $\mathsf{LL}_{\textrm{pol}}$ of linear logic, obtained by restriction  
to the polarised formulas:
$$P::= X\Alt P\otimes P \Alt P\oplus P \Alt !N\quad\quad\quad N::= X^\bot \Alt N\lpar N \Alt N\with N \Alt ?P$$
This fragment is also a fragment of  $\mathsf{LLP}$, (cf. \cite{LaurentTh}), up to the change of notation for the formulas: we write here $\overline{P}$ for $P^\bot$ and  
$\notP{\overline{N}}$ for $!N$.
Girards translation encodes call-by-name implication as follows:
 $$(A\rightarrow B)^*=!(A^*)\multimap B^*=(?(A^*)^\bot)\lpar B^*$$
 and then every 
 $\lambda$-term  $\Gamma\vdash M:A$ into a proof of $!(\Gamma)^*\vdash A^*$.
 Up to the inclusion of $\mathsf{LL}_{\textrm{pol}}$ into $\mathsf{LLP}$, up to the definition of the $\lbd$-calculus inside $\mathsf{LKT}$ given above, up to the change of notation, and up to the duality between $\mathsf{LKT}$ and $\mathsf{LKQ}$, Girard's translation coincides with the (restriction of) our translation above from $\mathsf{LKT}$ to $\mathsf{LLP}$.
 On the other hand, the restriction to the $\lambda\mu$-calculus of our translation from $\mathsf{LKT}$ to $\mathsf{NJ}$ is the CPS translation of Lafont-Reus-Streicher.
Thus, restricted to the $\lbd$-calculus, Proposition \ref{factor-cps} reads as follows:  
\begin{center}
{\em Lafont-Reus-Streicher's CPS factors through Girard's translation}.
\end{center}
Explicitly, on types, we have that $A^*$ coincides with $A$ as expanded in $\mathsf{LKT}$, and (cf. \cite{CH2000}),
starting from the simply-typed $\lambda$-term $(\Gamma\vdash M:A)$,
\begin{itemize}
\item we view $M$  as an expression $(\lkv{\Gamma}{}{M:A}{})$  of $\mathsf{LKT}$ (using the CBN encoding of implication),
\item and then as a context $(\lke{}{M:\overline{A}}{}{\overline{\Gamma}})$ of $\mathsf{LKQ}$,
\item then by our above translation we find back the result of Girard's translation 
$(\lke{\notP{(\overline{\Gamma})}}{M_{\mathsf{LLP}}:\overline{A}}{}{})$,
\item and we arrive finally at the Hofmann-Streicher CPS-transform  $(\notP{(\overline{\Gamma})}\vdash \CPSP{M}:\notP{(\overline{A})})$ of $M$, through the translation ${}_{\mathsf{NJ}}$.\footnote{The LRS translation of implication goes as follows:
$(A\rightarrow B)_{\textrm{LRS}}=
R^{A_{\textrm{LRS}}}\times B_{\textrm{LRS}}$, and we have 
$\CPSP{(\overline{A})}=A_{\textrm{LRS}}$
.}
\end{itemize}
But the above reading is actually stronger, because it is not hard to describe a translation ${}_{\mathsf{LJ}}$  inverse to ${}_{\mathsf{NJ}}$, so that up to this further isomorphism, we have that:
\begin{center}
{\em The LRS translation of the CBN $\lbd$-calculus coincides with Girard's translation}.
\end{center}
This nice story does not extend immediately to the $\lambda\mu$-calculus, for which the simplest extension of Girard's translation, taking $\lkv{\Gamma}{}{M:A}{\Delta}$ 
to a proof of $!(\Gamma^*)\vdash A^*,?(\Delta)^*$ is not polarised. In fact, Laurent \cite{LaurentTh} has shown that the natural target for an
extension of Girard's translation to CBN  $\lambda\mu$-calculus is $\mathsf{LLP}$, in which
we can spare the ?'s on the right, i.e., we can translate $\lkv{\Gamma}{}{M:A}{\Delta}$ into
a proof  of $!(\Gamma^*)\vdash A^*,\Delta^*$ (contractions on negative formulas are free in $\mathsf{LLP}$).  So, the extension of the picture to  call-by-name $\lbd\mu$-calculus is\footnote{See also \cite{LR2003} for further discussion.}:
\begin{center}
{\em The LRS translation of the CBN $\lbd\mu$-calculus coincides with Laurent-Girard's translation into $\mathsf{LLP}$}.
\end{center}

 \section{A synthetic system} \label{LKQS-sec}
 In this section we pursue two related goals.
 \begin{enumerate}
\item We want to account for the full (or strong) focalisation  (cf. \cite{QuaTor96}), which consists in removing the use of contractions in  the negative phases and carrying 
 these phases maximally, up to having only atoms on the left of the sequent. The positive phases are made also  ``more maximal''  by allowing the use of the axiom only on positive atoms $X$. This is of interest in a proof search perspective, since the stronger discipline
 further reduces the search space.
\item We would like our syntax to quotient proofs over the order of decomposition of negative formulas.  The use of  structured pattern-matching 
(cf. Examples \ref{LKQ-proofs-ex}, \ref{LKQ-exp-ex}) is relevant, as  we can describe the construction of a proof of
$(\Gamma, x:(P_1\otimes P_2)\otimes(P_3\otimes P_4)\vdash \Delta)$ out of a proof of
$c:(\Gamma, x_1:P_1,x_2:P_2,x_3:P_3,x_4:P_4\vdash\Delta)$ ``synthetically'', by writing
$\coupe{\Vtov{x}}{\tilde\mu((x_1,x_2),(x_3,x_4)).c}$, where $\tilde\mu((x_1,x_2),(x_3,x_4)).c$ stands for an abbreviation of either of the following two commands:
$$\begin{array}{lll}
\coupe{\Vtov{x}}{\tilde\mu(y,z).\coupe{\Vtov{y}}{\tilde\mu(x_1,x_2).\coupe{\Vtov{z}}{\tilde\mu(x_3,x_4).c}}} & \quad\quad & \coupe{\Vtov{x}}{\tilde\mu(y,z).\coupe{\Vtov{z}}{\tilde\mu(x_3,x_4).\coupe{\Vtov{y}}{\tilde\mu(x_1,x_2).c}}}
\end{array}$$
\end{enumerate}
The two goals are connected, since applying strong focalisation will forbid the formation of these
two terms (because $y,z$ are values appearing with non atomic types), keeping the synthetic form only...  provided we make it first class.  

\smallskip
We shall proceed in {\em two steps}. The first, intermediate one consists in introducing first-class {\em counterpatterns} and will serve goal 1 but not quite goal 2:
$$\begin{array}{lllllllll}
\mbox{Simple commands} && c::=\coupe{v}{e} &\quad\quad\quad\quad& \mbox{Commands} && C::= c\Alt \copaireC{C}{q}{q}{C}\\
\mbox{Expressions} && v::= \Vtov{V}\Alt \mu\alpha.C &\quad\quad\quad\quad&
\mbox{Values} && V::= x \Alt  (V,V) \Alt \inl{V}\Alt \inr{V}\Alt e^\bullet \\
\mbox{Contexts} && \fbox{$e::=\alpha \Alt \tilde\mu q.C$} &\quad\quad\quad\quad&
 \mbox{Counterpatterns} && \fbox{$q::= x \Alt \alpha^\bullet \Alt (q,q) \Alt\copaireP{q,q}$}
\end{array}$$
The counterpatterns  are to be thought of as  constructs that match patterns (see below).

In this syntax, we have gained a unique $\tilde\mu$ binder, but the price to pay (provisionally) is
that now commands are trees of copairings $\copaireC{\_}{q_1}{q_2}{\_}$ whose leaves are simple commands.

The typing discipline is restricted with respect to that of Figure 1 (and adapted to the setting with explicit counterpatterns). Let  $\Xi=x_1:X_1,\ldots,x_n:X_n$ denote a left context consisting of {\it atomic formulas only}.  The rules are as follows:
{\small $$\begin{array}{c}\seq{}{\lkV{\Xi\:,\:x:X}{}{x:X}{\Delta}}\quad\quad
\seq{}{\lke{\Xi}{\alpha:P}{}{\alpha:P\:,\:\Delta}}\quad\quad
\seq{\lkv{\Xi}{}{v:P}{\Delta}\quad\quad
\lke{\Xi}{e:P}{}{\Delta}}
{\coupe{v}{e}:(\lkc{\Xi}{}{}{\Delta})}
\\\\
\seq{C:(\lkc{\Xi\:,\:q:P}{}{}{\Delta})}{\lke{\Xi}{\tilde\mu q.C:P}{}{\Delta}}
\quad\quad \seq{C:(\lkc{\Xi}{}{}{\alpha:P\:,\:\Delta})}{\lkv{\Xi}{}{\mu\alpha.C:P}{\Delta}}\quad\quad \seq{\lkV{\Xi}{}{V:P}{\Delta}}{\lkv{\Xi}{}{\Vtov{V}:P}{\Delta}}
\\\\
\seq{\lke{\Xi}{e:P}{}{\Delta}}{\lkV{\Xi}{}{e^\bullet:\notP{P}}{\Delta}}\quad\quad\seq{\lkV{\Xi}{}{V_1:P_1}{\Delta} \quad\quad \lkV{\Xi}{}{V_2:P_2}{\Delta}}
{\lkV{\Xi}{}{(V_1,V_2):P_1\otimes P_2}{\Delta}}\quad\quad
\seq{\lkV{\Xi}{}{V_1:P_1}{\Delta}}{\lkV{\Xi}{}{\inl{V_1}:P_1\oplus P_2}{\Delta}}\quad\quad
\seq{\lkV{\Xi}{}{V_2:P_2}{\Delta}}{\lkV{\Xi}{}{\inr{V_2}:P_1\oplus P_2}{\Delta}}
\\\\\seq{C:(\lkc{\Gamma}{}{}{\alpha:P\:,\:\Delta})}{C:(\lkc{\Gamma\:,\:\alpha^\bullet:\notP{P}}{}{}{\Delta})}\quad\quad
\seq{C:(\lkc{\Gamma\:,\:q_1:P_1\:,\:q_2:P_2}{}{}{\Delta})}{C:(\lkc{\Gamma\:,\:(q_1,q_2):P_1\otimes P_2}{}{}{\Delta})}\quad\quad
\seq{C_1:(\lkc{\Gamma\:,\:q_1:P_1}{}{}{\Delta})\quad\quad C_2:(\lkc{\Gamma\:,\:q_2:P_2}{}{}{\Delta})}
{\copaireC{C_1}{q_1}{q_2}{C_2}:(\lkc{\Gamma\:,\:\copaireP{q_1,q_2}:P_1\oplus P_2}{}{}{\Delta})}
\end{array}$$}

Our aim now ({\em second step}) is to get rid of the tree structure of a command. Indeed, towards our second goal, if $c_{ij}:(\Gamma,x_i:P_i,x_j:P_j\vdash_S\Delta)$ ($i=1,2, j=3,4$), we want to identify
$\copaireC{\copaireC{c_{13}}{x_3}{x_4}{c_{14}}}{x_1}{x_2}{\copaireC{c_{23}}{x_3}{x_4}{c_{24}}}$
and 
$\copaireC{\copaireC{c_{13}}{x_1}{x_2}{c_{23}}}{x_3}{x_4}{\copaireC{c_{14}}{x_1}{x_2}{c_{24}}}
$.
To this effect, we need a last ingredient.
We introduce a syntax of {\it patterns}, and we redefine the syntax of values, as follows:
\begin{center}
\fbox{${\cal V}::=x\Alt e^\bullet\quad\quad\quad
V::=p\:\patc{\Subi{i}{{\cal V}_i}}{i\in p}
\quad\quad\quad\quad p::= x \Alt \alpha^\bullet \Alt (p,p) \Alt\inl{p} \Alt\inr{p}$}
\end{center}
where
$i\in p$ is defined by:
$$\seq{}{x\in x}\quad\seq{}{\alpha^\bullet\in\alpha^\bullet}\quad
\seq{i\in p_1}{i\in (p_1,p_2)}\quad\seq{i\in p_2}{i\in(p_1,p_2)}\quad
\seq{i\in p_1}{i\in\inl{p_1}}\quad\seq{i\in p_2}{i\in \inr{p_2}}$$
Moreover, ${\cal V}_i$  must be of the form $y$ (resp. $e^\bullet$) if $i=x$ (resp.
$i=\alpha^\bullet$).

Patterns are required to be linear, as well as the  counterpatterns, for which the definition of ``linear'' is adjusted in the case $\copaireP{q_1,q_2}$, in which a variable can occur (but recursively linearly so) in both $q_1$ and $q_2$.

Note also that the reformulation of values is up to $\alpha$-conversion:  for example, it is understood that $\alpha^\bullet\langle\Subi{\alpha^\bullet}{e^\bullet}\rangle = \beta^\bullet\langle\Subi{\beta^\bullet}{e^\bullet}\rangle$

\medskip
We can now  rephrase the logical reduction rules in terms of pattern/counterpattern interaction (whence
the terminology), resulting in the following packaging of rules:
\begin{center}
$\seq{V=p\:\langle\ldots \Subi{x}{y},\ldots,\Subi{\alpha^\bullet}{e^\bullet},\ldots\rangle\quad\quad\Sub{C}{q}{p}\longrightarrow^* c}{ \coupe{\Vtov{V}}{\tilde\mu q.C} \longrightarrow \Submimp{c}{\ldots, \Subi{x}{y},\ldots,\Subi{\alpha}{e},\ldots}
}$
\end{center}
where $c\{\sigma\}$ is the usual, implicit substitution, and where $c$ (see the next proposition) is the normal form of 
$\Sub{C}{q}{p}$  with respect to the following set of rules:
 $$\begin{array}{l}
\Subm{C}{\Subi{(q_1,q_2)}{(p_1,p_2)},\sigma}\longrightarrow
\Subm{C}{\Subi{q_1}{p_1},\Subi{q_2}{p_2},\sigma}\\
\Subm{\copaireC{C_1}{q_1}{q_2}{C_2}}{\Subi{\copaireP{q_1,q_2}}{\inl{p_1}},\sigma}\longrightarrow
\Subm{C_1}{\Subi{q_1}{p_1},\sigma}\quad\quad
\Subm{\copaireC{C_1}{q_1}{q_2}{C_2}}{\Subi{\copaireP{q_1,q_2}}{\inr{p_2}},\sigma}\longrightarrow
\Subm{C_2}{\Subi{q_2}{p_2},\sigma}\\
\Subm{C}{\Subi{\alpha^\bullet}{\alpha^\bullet},\sigma}\longrightarrow
\Subm{C}{\sigma}\quad\quad\quad
\Subm{C}{\Subi{x}{x},\sigma}\longrightarrow
\Subm{C}{\sigma}\\
\end{array}$$
Logically, this means that we now consider each formula as made of blocks of {\em synthetic} connectives. 
\begin{example} 
\begin{itemize}
\item Patterns for $P=X\otimes(Y\oplus\notP{Q})$.
Focusing on the right yields two possible proof searches:
$$\seq{\lkV{\Gamma}{}{x'\set{{\cal V}_{x'}}:X}{\Delta}
\quad\lkV{\Gamma}{}{y'\set{{\cal V}_{y'}}:Y}{\Delta}}{\lkV{\Gamma}{}{{ (x',\inl{y'})}\set{{\cal V}_{x'},{\cal V}_{y'}}:X\otimes(Y\oplus\notP{Q})}{\Delta}}
\quad\quad
\seq{\lkV{\Gamma}{}{x'\set{{\cal V}_{x'}}:X}{\Delta}\quad\lkV{\Gamma}{}{{\alpha'}^\bullet\set{{\cal V}_{{\alpha'}^\bullet}}:\notP{Q}}{\Delta}}{\lkV{\Gamma}{}{{ (x',\inr{{\alpha'}^\bullet})}\set{{\cal V}_{x'},{\cal V}_{{\alpha'}^\bullet}}:X\otimes(Y\oplus\notP{Q})}{\Delta}}$$
\item Counterpattern for $P=X\otimes(Y\oplus\notP{Q})$.
The counterpattern describes the tree structure of $P$:
$$\seq{{ c_1:(\lkc{\Gamma\:,\:x:X\:,\:y:Y}{}{}{\Delta})}\quad\quad { c_2:(\lkc{\Gamma\:,\:x:X\:,\:\alpha^\bullet:\notP{Q}}{}{}{\Delta})}}
{\copaireC{c_1}{y}{\alpha^\bullet}{c_2}:(\lkc{\Gamma\:,\:{(x,\copaireP{y,\alpha^\bullet})}:X\otimes(Y\oplus\notP{Q})}{}{}{\Delta})}$$
\end{itemize}
We observe that the { leaves} of the decomposition are in  one-to-one correspondence with the patterns
$p$ for the (irreversible) decomposition  of $P$ on the right:
$$\begin{array}{l}\Sub{\copaireC{c_1}{y}{\alpha^\bullet}{c_2}}{{ q}}{{ p_1}} \longrightarrow^* c_1\quad\quad
\Sub{\copaireC{c_1}{y}{\alpha^\bullet}{c_2}}{{ q}}{{ p_2}}\longrightarrow^* c_2
\end{array}$$
where ${ q=(x,\copaireP{y,\alpha^\bullet})}\;, \;{ p_1=(x,\inl{y})}\;, \;{ p_2=(x,\inr{\alpha^\bullet})}$.
\end{example}
This correspondence is general. We define two predicates
$c\in C$ and $\suits{q}{p}$  (``$q$ is orthogonal to $p$'')  as follows:
$$\seq{}{c\in c}\quad\quad\seq{c\in C_1}{c\in\copaireC{C_1}{q_1}{q_2}{C_2}}\quad\quad
\seq{c\in C_2}{c\in\copaireC{C_1}{q_1}{q_2}{C_2}}$$
\begin{center}
\fbox{$\seq{}{\suits{x}{x}} \quad\quad \seq{}{\suits{\alpha^\bullet}{\alpha^\bullet}}\quad\quad 
\seq{\suits{q_1}{p_1}\;\;\suits{q_2}{p_2}}{\suits{(q_1,q_2)}{(p_1,p_2)}}\quad\quad
\seq{\suits{q_1}{p_1}}{\suits{\copaireP{q_1,q_2}}{\inl{p_1}}}\quad\quad \seq{\suits{q_2}{p_2}}{\suits{\copaireP{q_1,q_2}}{\inr{p_2}}} \quad\quad\quad$}
\end{center} 

\medskip
We can now state the correspondence result.
\begin{proposition} \label{pattern-counterpatten-corr}
Let 
{ $C:(\lkc{\Xi\,,\,q:P}{}{}{\Delta})$} (as in the assumption of the typing rule for $\tilde\mu q.C$), and let $p$ be such that $q$ is orthogonal to $p$.
Then the normal form $c$ of $\Sub{C}{q}{p}$ is a simple command, and the mapping $p\mapsto c$ ($q,C$ fixed) from  
$\setc{p}{\suits{q}{p}}$ to $\setc{c}{c\in C}$  is one-to-one and onto.
\end{proposition}
\Proof The typing and the definition of orthogonality entail that  in all intermediate $\Subm{C}{\sigma}$'s the substitution $\sigma$ has an item for each counterpattern in the sequent, and that reduction progresses. 
The rest is easy.  (Note that a more general statement is needed for the induction to go through, replacing $\Xi\,,\,q:P$, ``$q$ orthogonal to $p$'', and $\Sub{C}{q}{p}$ with
$\Xi\,,\,q_1:P_1\,,\,\ldots\,,\,q_n:P_n$, ``$q_i$ orthogonal to $p_i$ for $i=1,\ldots,n$'', and $\Subm{C}{\Subi{q_1}{p_1},\ldots\Subi{q_n}{p_n}}$, respectively.) \qed

\begin{figure}[t] \label{big-step-mu-tilde}
\caption{The syntax and reduction semantics of $\mathsf{L}_{\textrm{synth}}$}
$$\begin{array}{lllll}
c::=\coupe{v}{e} \quad\quad\quad\quad v::=\Vtov{V}\Alt \mu\alpha.c\\
V::=p\:\patc{\Subi{i}{{\cal V}_i}}{i\in p} \quad\quad {\cal V}::=x\Alt e^\bullet &&&& p::= x \Alt \alpha^\bullet \Alt (p,p) \Alt\inl{p} \Alt\inr{p}\\
e::=\alpha\Alt  \tilde\mu q.\setc{p\mapsto c_p}{\suits{q}{}{p}} &&&& q::= x \Alt \alpha^\bullet \Alt (q,q) \Alt\copaireP{q,q}
\end{array}$$
$$\begin{array}{|cc|c|}
\hline
(\tilde\mu^{\scriptscriptstyle +}) \;\; \coupe{\Vtov{(p\:\langle\dots,\Subi{x}{y},\ldots,\Subi{\alpha^\bullet}{e^\bullet}\ldots\rangle)}}{\tilde\mu q.\setc{p\mapsto c_p}{\suits{q}{}{p}}}  \longrightarrow  {c_p}\,\set{\ldots,\Subi{x}{y},\ldots,\Subi{\alpha}{e},\ldots\rangle} 
&&\;\;  (\mu)\;\; \coupe{\mu\alpha.c}{e}  \longrightarrow \Subimpl{c}{\alpha}{e}\\\hline
\end{array}$$

Typing rules: the old ones for $\alpha,x,e^\bullet,c$, plus the following ones:
$$\seq{\ldots \quad\quad\lkV{\Gamma}{}{{\cal V}_i:P_i}{\Delta}\quad((i:P_i)\in\Gamma(p,P))\quad\quad\ldots}{\lkV{\Gamma}{}{p\:\patc{\Subi{i}{{\cal V}_i}}{i\in p}:P}{\Delta}}\quad\quad
\seq{\ldots \quad\quad c_p:(\lkc{\Gamma\,,\,\Xi(p,P)}{}{}{\Delta(p,P)\,,\,\Delta})\quad(\suits{q}{p}) \quad\quad\ldots}{\lke{\Gamma}{\tilde\mu q.\setc{p\mapsto c_p}{\suits{q}{}{p}}:P}{}{\Delta}}$$
where $\Gamma(p,P)$ must be successfully defined as follows:
$$\begin{array}{l}
\Gamma(x,X)=(x:X)\quad\quad
\Gamma(\alpha^\bullet,\notP{P})=(\alpha^\bullet:\notP{P})\\
\Gamma((p_1,p_2),P_1\otimes P_2)=\Gamma(p_1,P_1)\,,\,\Gamma(p_2,P_2)\quad\quad
\Gamma(\inl{p_1},P_1\oplus P_2)= \Gamma(p_1,P_1)\quad\quad
\Gamma(\inr{p_2},P_1\oplus P_2)= \Gamma(p_2,P_2)
\end{array}$$
and where
$$
\Xi(p,P)=\setc{x:X}{x:X\in\Gamma(p,P)} \quad\quad \Delta(p,P)=\setc{a:P}{\alpha^\bullet:\notP{P}\in\Gamma(p,P)}$$
\Figbar
\end{figure}

\medskip
Thanks to this correspondence, we can quotient over the ``bureaucracy'' of commands, and we arrive at the calculus
described in Figure 4, together with its typing rules,
which we call {\em synthetic system }$\mathsf{L}$, or $\mathsf{L}_{\textrm{synth}}$. 
The $\tilde\mu$ construct of $\mathsf{L}_{\textrm{synth}}$ is closely related to Zeilberger's higher-order abstract approach to focalisation in \cite{ZeilbergerCU}: indeed we can view
$\setc{p\mapsto c}{\suits{q}{}{p}}$ as a function from patterns to commands. We actually prefer to see here a {\em finite} record whoses fields are the $p$'s orthogonal to $q$. 
There are only two reduction rules in $\mathsf{L}_{\textrm{synth}}$. the $\mu$-rule now expressed with implicit substitution and the $\tilde\mu^+$-rule, which
combines two familiar operations: select a field $p$ (like in object-oriented programming), and substitute (like in  functional programming). The next proposition relates $\mathsf{L}_{\textrm{synth}}$  to $\mathsf{L}_{\textrm{foc}}$.
\begin{proposition} \label{synth-foc-comp}
The typing system of   $\:\mathsf{L}_{\textrm{synth}}$ is complete\footnote{It is also easy to see that the underlying translation from $\mathsf{L}_{\textrm{foc}}$ to $\mathsf{L}_{\textrm{synth}}$ is reduction-reflecting (cf. Remark \ref{LK-LKQ-not-red-refl}).} with respect to  $\LKQ$.
\end{proposition}

\Proof  The completeness of $\mathsf{L}_{\textrm{synth}}$ with respect to the intermediate system above is an easy consequence of
Proposition \ref{pattern-counterpatten-corr}.  We are thus left with proving the completeness of the intermediate system.
We define a rewriting relation between sets of sequents as follows:
\begin{center}
$\begin{array}{l}
(\lkc{\Gamma,x:\notP{P}}{}{}{\Delta}),{\bf S} \rightsquigarrow
(\lkc{\Gamma}{}{}{\alpha: P,\Delta}),{\bf S} 
\\
(\lkc{\Gamma,x:P_1\otimes P_2}{}{}{\Delta}),{\bf S} \rightsquigarrow
(\lkc{\Gamma,x_1:P_1,x_2:P_2}{}{}{\Delta}),{\bf S}
\\
(\lkc{\Gamma,x:P_1\oplus P_2}{}{}{\Delta}),{\bf S}
\rightsquigarrow
(\lkc{\Gamma,x_1:P_1}{}{}{\Delta}),(\lkc{\Gamma,x_2:P_2}{}{}{\Delta}),{\bf S}
\end{array}$
\end{center}
 (where $\alpha,x_1,x_2$ are fresh).
A normal form for this notion of reduction is clearly a set of sequents of the form $\lkc{\Xi}{}{}{\Delta}$. It is also easy to see that
 $\rightsquigarrow$ is confluent and (strongly) normalising. 
In what follows, $\vdash_S$ (resp. $\vdash$) will signal the  intermediate proof system (resp. $\mathsf{L}_{\textrm{foc}}$). The following property is easy to check.

\medskip
If  $(\Xi_1\vdash \Delta_1),\ldots,(\Xi_n\vdash\Delta_n)$ is the normal form of 
$(x_1:P_1,\ldots,x_m:P_m\vdash\Delta)$ for $\rightsquigarrow$ and if 
$c_i:(\Xi_i\vdash_S\Delta_i)$, 
then there exist $q_1,\ldots,q_m$ and a command 
$C:(q_1:P_1,\ldots,q_m:P_m\vdash_S\Delta)$
whose leaves are the $c_i$'s.

\smallskip
We prove the following properties together:

\smallskip\noindent
1) If $c:(x_1:P_1,\ldots,x_m:P_m\vdash\Delta)$, then there exist $q_1,\ldots,q_m$ and $C$ such that $C:(q_1:P_1,\ldots,q_m:P_m\vdash_S\Delta)$.

\noindent
2) If $\lke{\Xi}{e:P}{}{\Delta}$, then there exists $e'$ such that $\lkes{\Xi}{e':P}{}{\Delta}$ (and similarly for expressions $v$).

\smallskip
The proof is by induction on a notion of size which is the usual one except that the size of a variable $x$ is not 1 but the size of its type.  It is easy to check that the substitutions involved in Lemma \ref{e-subst-lemma} do not increase the size. The interesting case is $c=\coupe{v}{e}$.
Let $(\Xi_1\vdash \Delta_1),\ldots,(\Xi_n\vdash\Delta_n)$ be the normal form of 
$(x_1:P_1,\ldots,x_m:P_m\vdash\Delta)$. Then, by repeated application of Lemma \ref{e-subst-lemma}, we get  $v_1,\ldots,v_n$ and $e_1,\ldots,e_n$, and then by induction $v'_1,\ldots,v'_n$ and $e'_1,\ldots,e'_n$ which we assemble pairwise to form $\coupe{v'_1}{e'_1},\ldots,
\coupe{v'_n}{e'_n}$, which  in turn, as noted above, can be assembled in a tree $C:(q_1:P_1,\ldots,q_m:P_m\vdash_S\Delta)$. \qed

\smallskip Putting together Propositions \ref{LKQ-complete} and \ref{synth-foc-comp}, we have proved that $\mathsf{L}_{\textrm{synth}}$ is complete with respect to $\LK$ for provability.

\begin{remark} \label{Boehm-ludique}
 \begin{itemize}
\item In the multiplicative case (no $C$, $\inl{V}$, $\inr{V}$, $\copaireP{q_1,q_2}$),
there is a unique $p$ such that $\suits{q}{}{p}$, namely $q$, and the syntax boils down to
\begin{center}
$
{\cal V}::=x \Alt e^\bullet\quad
 V::=p\:\patc{\Subi{i}{{\cal V}_i}}{i\in p}\quad
{ v::=x\Alt \tilde\mu q.\set{c}}\quad
{ c::=\coupe{\Vtov{V}}{\alpha}}
$
\end{center}
Compare with   {\it B\"ohm trees}:
$\begin{array}{lll}
M::=\overbrace{\lbd\vec{x}.\underbrace{P}_{{ c}}}^{{ e}} &\quad\quad&
P::=y\underbrace{\overbrace{M_1}^{{ {\cal V}}}\ldots \overbrace{M_n}^{{ {\cal V}}}}_{{ V}}
\end{array}$.
For example (cf. the CBN translation in Section \ref{encodings-sec}),
if $M_j$ translates to $e_j$, then
$\lbd x_1x_2.xM_1M_2M_3$ translates to 
$\tilde\mu(\tilde{x_1}^\bullet,\tilde{x_2}^\bullet,y).\coupe{\Vtov{p\patc{\Subi{i}{{\cal V}_i}}{i\in p}}}{\tilde{x}}$, 
where $p=(\alpha_1^\bullet,\alpha_2^\bullet,\alpha_3^\bullet,z)$, ${\cal V}_{z_j}=(e_j)^\bullet$
, and ${\cal V}_z=y$.
\item (for readers familiar with \cite{GirardLS})
 Compare with the syntax for ludics presented in \cite{CuLLintroII}:

$\begin{array}{l}
\overbrace{M}^{{ e}}::=  \setc{\overbrace{J}^p \mapsto \lbd\setc{x_j}{j\in J}.P_J}{\overbrace{J\in{\cal N}}^{\suits{q}{}{p}}}\quad\quad
\underbrace{P}_{ c}::=(x\cdot \underbrace{\overbrace{I}^p)\setc{M_i}{i\in I}}_{ V}\Alt \Omega \Alt \maltese
\end{array}$

\end{itemize}
\end{remark}

 \section{Conclusion} \label{conclusion-sec}
 We believe that  Curien-Herbelin's syntactic kit, which we could call {\em system $\mathsf{L}$} for short,  provides us with a robust infrastructure for  proof-theoretical investigations, and for applications in 
formal studies in operational semantics. Thus, the work presented here  is faithful to the spirit of Herbelin's Habilitation Thesis \cite{HerbelinHabil}, where he advocated an incremental approach to connectives, starting from a pure control kernel.


On the proof-theoretical side, we note, with respect to the original setting of ludics  \cite{GirardLS}, that a pattern $p$ is more precise than a ramification $I$ (finite  tree of subaddresses vs a set of immediate subaddresses). We might use this additional precision to design a version of ludics where axioms are first-class rather than treated as  infinite expansions.

On the side of applications to programming language semantics, the good fit between abstract machines and our  syntax $\mathsf{L}_{\textrm{foc}}$ makes it a good candidate for being used as an intermediate language appropriate to reason about the correctness of abstract machines (see also \cite{Levy2004}).
In this spirit, in order to account for languages with mixed call-by-value / call-by-name features, one may give a truly bilateral presentation of $\mathsf{L}_{\textrm{foc}}$ that  freely mixes positive and negative formulas like in Girard's $\mathsf{LC}$ \cite{GirardLC}.\footnote{See also \cite{MurthyLC} for an early analysis of the computational meaning of $\mathsf{LC}$ from a programming language perspective.} Such a system is presented in the long version of \cite{Munch2009}.

Finally, we wish to thank Bob Harper, Hugo Herbelin, and Olivier Laurent for helpful discussions.

 \end{document}